\documentclass[fleqn,review,3p,times]{elsarticle}

\usepackage{amssymb}
\usepackage{amsthm}

\usepackage{amsmath}
\usepackage{mathtools}
\usepackage{mathdots}
\usepackage{subcaption}

\usepackage{shellesc}
\usepackage{tikz,tikzscale}
\usetikzlibrary{positioning,patterns,calc}
\usetikzlibrary{external}

\usepackage{pgfplots}
\usepgfplotslibrary{statistics}
\usepgfplotslibrary{fillbetween}
\usepgfplotslibrary{colormaps}
\pgfplotsset{compat=1.13}
\pgfplotsset{lua backend=true}

\usepackage{tabularx}

\graphicspath{{fig/}}

\usepackage{xcolor}

\usepackage[hidelinks]{hyperref}

\mathchardef\mhyphen="2D

\newcommand{\stat}[1]{#1}
\newcommand{\mov}[1]{\tilde{#1}}
\renewcommand{\vec}[1]{\underline{#1}}
\newcommand{\co}{\eta}
\newcommand{\sco}{\stat{\co}}
\newcommand{\mco}{\mov{\co}}
\newcommand{\si}{\stat{i}}
\newcommand{\mi}{\mov{i}}
\newcommand{\mcopa}{\mco_{\parallel}}

\newcommand{\scopa}{\sco_{\parallel}}
\newcommand{\scope}{\sco_{\perp}}
\newcommand{\mipa}{\mi_{\parallel}}

\newcommand{\sipa}{\si_{\parallel}}
\newcommand{\sipe}{\si_{\perp}}
\newcommand{\lpa}{l_{\parallel}}

\newcommand{\Z}{\mathbb{Z}}
\newcommand{\disp}{{\Delta}}
\newcommand{\im}{i_\text{sub}}
\newcommand{\face}{\Omega}

\newcommand{\rank}{r}
\newcommand{\srank}{\stat{\rank}}
\newcommand{\mrank}{\mov{\rank}}
\newcommand{\flexiSideID}{{i_{\text{face}}}}
\newcommand{\iMortar}{i_\text{mortar}}
\newcommand{\mapping}{m}
\newcommand{\sortedArray}{A}

\newcommand{\figref}[1]{Figure~\ref{#1}}
\newcommand{\tabref}[1]{Table~\ref{#1}}
\newcommand{\secref}[1]{Section~\ref{#1}}
\LetLtxMacro{\originaleqref}{\eqref}
\renewcommand{\eqref}{Eq.~\originaleqref}

\newcommand{\toprule}{\hline\noalign{\vskip 2pt}}
\newcommand{\midrule}{\noalign{\vskip 1pt}\hline\noalign{\vskip 2pt}}
\newcommand{\bottomrule}{\noalign{\vskip 1pt}\hline}
\newcommand{\norule}{\noalign{\vskip 4pt}}

\newcommand{\ppvector}[1]{\ensuremath{\vec{#1}}}
\newcommand{\ppmatrix}[1]{\ensuremath{\underline{#1}}}

\newcommand{\unit}[1]{\ensuremath{\:\mathrm{#1}}}

\newcommand{\phyflux}{\ensuremath{F}}
\newcommand{\flux}{\ensuremath{\mathcal{F}}}

\newcommand{\inviscid}[1]{\ensuremath{#1^{c}}}
\newcommand{\viscous}[1]{\ensuremath{#1^{v}}}

\newcommand{\jacobi}{\ensuremath{\mathcal{J}}}

\newcommand{\element}{\ensuremath{E}}

\newcommand{\hazelhen}{\textit{Hazel Hen}}

\journal{}

\begin{document}

\begin{frontmatter}

\title{An Efficient Sliding Mesh Interface Method for High-Order Discontinuous Galerkin Schemes}

\author[label1]{Jakob D\"urrw\"achter\fnref{fn1}\corref{cor1}}
\ead{jd@iag.uni-stuttgart.de}

\author[label1]{Marius Kurz\fnref{fn1}}
\ead{m.kurz@iag.uni-stuttgart.de}

\author[label2]{Patrick Kopper}
\ead{kopper@ila.uni-stuttgart.de}

\author[label1]{Daniel Kempf}
\ead{kempf@iag.uni-stuttgart.de}

\author[label1]{Claus-Dieter Munz}
\ead{munz@iag.uni-stuttgart.de}

\author[label1]{Andrea Beck}
\ead{beck@iag.uni-stuttgart.de}

\fntext[fn1]{J. D\"urrw\"achter and M. Kurz share first authorship.}
\address[label1]{Institute of Aerodynamics and Gas Dynamics, University of Stuttgart, Pfaffenwaldring 21, 70569 Stuttgart, Germany}
\address[label2]{Institute of Aircraft Propulsion Systems, University of Stuttgart, Pfaffenwaldring 6, 70569 Stuttgart, Germany}

\cortext[cor1]{Corresponding author}

\begin{abstract}
Sliding meshes are a powerful method to treat deformed domains in computational fluid dynamics, where different parts of the domain are in relative motion. In this paper, we present an efficient implementation of a sliding mesh method into a discontinuous Galerkin compressible Navier-Stokes solver and its application to a large eddy simulation of a 1-1/2 stage turbine.
The method is based on the mortar method and is high-order accurate. It can handle three-dimensional sliding mesh interfaces with various interface shapes. For plane interfaces, which are the most common case, conservativity and free-stream preservation are ensured. We put an emphasis on efficient parallel implementation. Our implementation generates little computational and storage overhead. Inter-node communication via MPI in a dynamically changing mesh topology is reduced to a bare minimum by ensuring a priori information about communication partners and data sorting. We provide performance and scaling results showing the capability of the implementation strategy.
Apart from analytical validation computations and convergence results, we present a wall-resolved implicit LES of the 1-1/2 stage Aachen turbine test case as a large scale practical application example.
\end{abstract}

\begin{keyword}
Sliding mesh \sep Discontinuous Galerkin \sep High-order methods \sep High-performance computing \sep Large eddy simulation \sep Turbine flow
\end{keyword}

\end{frontmatter}

\section{Introduction}
\label{sec:introduction}
\begin{figure}[htb]
	\centering
	\begin{subfigure}{0.32\textwidth}
		\centering
		\includegraphics[width=0.6\textwidth]{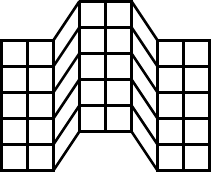}
	\end{subfigure}
	\begin{subfigure}{0.32\textwidth}
		\centering
		\includegraphics[width=0.6\textwidth]{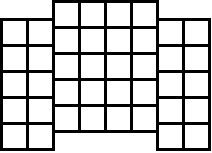}
	\end{subfigure}
	\begin{subfigure}{0.32\textwidth}
		\centering
		\includegraphics[width=0.6\textwidth]{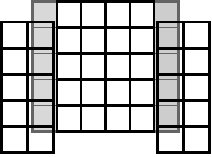}
	\end{subfigure}
	\caption{Schematics of common moving mesh techniques: the arbitrary~Lagrangian-Eulerian~approach~(ALE)~(\textit{left}), the~sliding~mesh~method (\textit{middle}) and the overset~mesh or Chimera method (\textit{right})}
	\label{fig:MovingMesh}
\end{figure}

High-order methods have significantly gained popularity over the last decade, due to their potential advantages in terms of accuracy and efficiency for many applications~\cite{wang2013}. A variety of high-order, element-based schemes has been proposed in literature, such as reconstruction-based approaches like the WENO schemes~\cite{liu1994,balsara2000}, h/p finite element schemes~\cite{nek5000-web-page,cantwell2015nektar++}, the flux reconstruction method~\cite{huynh2007} and the spectral difference method \cite{liu2006,wang2007}, with the latter two also being closely related to the discontinuous Galerkin methods \cite{may2011,degrazia2014}. In particular the discontinuous Galerkin spectral element method (DGSEM) \cite{hesthaven2007,hindenlang2012} has shown its suitability for simulations of unsteady turbulent flows in large-scale applications and high-performance computing (HPC) as consequence of its high-order of accuracy as well as its excellent scaling properties~\cite{wang2013,atak2016,krais2020flexi,beck2014,bolemann2015high}. While the development of novel formulations and efficient schemes is still ongoing, there is sufficient evidence from published results that DG and related methods have reached a certain level of maturity. Thus, adding and exploring new features to the existing formulations - while retaining the original properties - has now become another focus of ongoing development.\\
Many fields of engineering interest, as e.g. turbomachinery, wind turbines and rotorcraft, are characterized by moving geometries and large periodic displacements. A common technique to incorporate this movement into numerical schemes is the arbitrary Lagrangian-Eulerian (ALE) approach~\cite{minoli2011}, which introduces a time-dependent mapping from an arbitrarily deformed domain to the undeformed reference. Since the mesh connectivity remains unchanged, this approach is often limited to small relative displacements to retain valid grid cells.
In order to accommodate larger displacements, two approaches are commonly employed: In the overset mesh (Chimera) method \cite{wang2014,ahmad1996,pomin2002,sankaran2011,zahle2009}, several independent meshes inside the computational domain are overlaid, which can be moved independently and are coupled by means of the overlapping elements and associated inter-mesh interpolation operators. In the sliding mesh method~\cite{van2006,bakker1997,jaworski1997,ng1998,mcnaughton2014}, the computational domain is divided into non-overlapping sub-domains, which can slide along a common interface while preserving the mesh geometry inside each sub-domain. This approach can be seen as a special case of the Chimera method, where the overlap area is restricted to a linear (in 2D) or planar (in 3D) shared region, and the movement is prescribed accordingly. For both approaches, special care must be taken to construct high-order variants, which guarantee global conservation and overall error scaling behaviour, especially if the consistent geometry representation of curved interfaces has to be taken into account~\cite{zhang2015,zhang2016,qiu2019,ferrer2012,ramirez2015,wurst2015,brazell2016}.
Schematics of the discussed methods are shown in~\figref{fig:MovingMesh}.\\
While the overset mesh method allows greater flexibility in both kinematics and interface topology, the sliding mesh approach is conceptually simpler and more efficient for a certain class of problems, e.g. for rotor-stator interactions. In this work, we thus focus on the challenge of combining such an approach with a high-order discontinuous Galerkin solver. This not only requires a conservative, high-order accurate method at the sub-domain interface, but also the design of an efficient and dynamic parallelization strategy which retains the scaling of the static version. These efforts then give us the ability to conduct high-order sliding mesh simulations at an industrial scale, and provide a novel tool for investigating the highly non-linear and unsteady flow physics in these scenarios.
The implementation considered in the present work is open source and can be retrieved from GitHub\footnote{\texttt{https://github.com/flexi-framework/flexi-extensions/tree/sliding-mesh}}.
\\
The outline of this paper is as follows: The governing equations on the moving domain are given in~\secref{sec:equations} and the DGSEM scheme as well as the treatment of non-conforming meshes are discussed in~\secref{sec:numerics}. \secref{sec:implementation} gives details on the proposed implementation and parallelization strategy for the sliding mesh method. In~\secref{sec:results}, we demonstrate the error convergence of the method, its scaling properties and the suitability for large-scale applications. To this end, an implicit, wall-resolved large eddy simulation (LES) of the 1-1/2 stage turbine test case~\textit{Aachen Turbine} is presented and discussed. \secref{sec:conclusion} concludes the paper, and we give an outlook on further developments.

\section{Governing Equations}
\label{sec:equations}
In this work, we consider the three-dimensional compressible Navier-Stokes equations, which can be written in conservative form as
\begin{equation}
	\ppvector{U}_t + \ppvector{\nabla}_x \cdot \inviscid{\ppvector{\phyflux}}(\ppvector{U}) - \ppvector{\nabla}_x \cdot \viscous{\ppvector{\phyflux}}\left(\ppvector{U},\ppvector{\nabla}_x \ppvector{U} \right)= 0 \;,
  \label{eq:navier_stokes}
\end{equation}
with the vector of conserved variables in usual notation $\ppvector{U} = \left[\rho,\rho v_1,\rho v_2,\rho v_3, \rho e\right]^T$ and its time derivative $\ppvector{U}_t$, the convective fluxes \inviscid{\ppvector{\phyflux}}, the viscous fluxes \viscous{\ppvector{\phyflux}} and the differential operator $\ppvector{\nabla}_x$ with respect to the physical coordinates $\ppvector{x}=\left[x_1,x_2,x_3\right]^T$.
To account for the moving frame of reference in case of mesh movement, the fluxes can be written in an arbitrary Lagrangian-Eulerian (ALE) formulation~\cite{hirt1974}, which yields the fluxes
with columns $i=1,2,3$ as
\begin{equation}
  \inviscid{\ppvector{\phyflux}}_i =
  \left[
  \begin{array}{c}
    \rho v_i \\
    \rho v_1 v_i + \delta_{1i} p\\
    \rho v_2 v_i + \delta_{2i} p\\
    \rho v_3 v_i + \delta_{3i} p\\
    \rho e v_i + p v_i
  \end{array}
  \right]
  - v^g_i
  \left[
  \begin{array}{c}
    \rho    \\
    \rho v_1\\
    \rho v_2\\
    \rho v_3\\
    \rho e
  \end{array}
  \right]
  \;,\;
  \viscous{\ppvector{\phyflux}}_i =
  \left[
  \begin{array}{c}
    0 \\
    \tau_{1i}\\
    \tau_{2i}\\
    \tau_{3i}\\
    \tau_{ij}v_j-q_i
  \end{array}
  \right] \;\;.
  \label{eq:fluxes_written_out}
\end{equation}
Here, $\ppvector{v}^g = \left[v^g_1,v_2^g,v^g_3\right]^T$ denotes the velocity of the grid in Cartesian coordinates, which induces an additional flux contribution, and $\delta$ denotes the Kronecker delta. The stress tensor~$\tau_{ij}$ and the heat flux~$q_i$ can be written as
\begin{align}
  \tau_{ij} &= \mu\left(\frac{\partial v_i}{\partial x_j}+\frac{\partial v_j}{\partial x_i}\right)+\lambda\,\delta_{ij}\frac{\partial v_k}{\partial x_k} \;\;,
  \label{eq:stress_tensor}\\
  q_i &= - k\:\frac{\partial T}{\partial x_i}    \;\;,
  \label{eq:heat_flux}
\end{align}
where $k$ denotes the heat conductivity and $T$ the static temperature. Moreover, Stokes' hypothesis states $\lambda = -\frac{2}{3}\mu$ with $\mu$ as dynamic viscosity. The equation system is closed with the perfect gas assumption which yields the equation of state
\begin{equation}
 p = \rho R T = \rho\left(\gamma-1\right) \left[e-\frac{1}{2}\left(v_1^2+v_2^2+v_3^2\right)\right]
\end{equation}
with the ratio of specific heats $\gamma$ and the specific gas constant $R$.\\

\section{Numerical Methods}
\label{sec:numerics}

\subsection{Discontinuous Galerkin Spectral Element Method}
\label{sec:DGSEM}
For the discontinuous Galerkin spectral element method (DGSEM), the computational domain is subdivided into non-overlapping elements. Each element is mapped into the reference element~$E\in\left[-1,1\right]^3$ using a polynomial mapping~$\ppvector{x}=\ppvector{x}\left(\ppvector{\xi}\right)$ as described in more detail in~\cite{kopriva2009}. By defining the Jacobian of the mapping~$\jacobi = \mathrm{det}\left(\partial x_i/\partial \xi_j\right)$ and introducing the contravariant fluxes~$\ppvector{\flux}$ (again, see \cite{kopriva2009} for details), \eqref{eq:navier_stokes} can be written in the reference space as
\begin{equation}
  \jacobi\left(\ppvector{\xi}\right) \ppvector{U}_t + \ppvector{\nabla}_{\xi} \cdot \ppvector{\flux}\left(\ppvector{U},\ppvector{\nabla}_{\xi}\ppvector{U}\right) = \ppvector{0} \;.
  \label{eq:transformed_equation}
\end{equation}
The projection of \eqref{eq:transformed_equation} onto the polynomial test space, spanned by test functions $\phi\left(\ppvector{\xi}\right)$ in the reference element \element, and subsequent integration by parts yields the weak formulation as
\begin{equation}
  \int_{\element} \jacobi\left(\ppvector{\xi}\right)\ppvector{U}_t\phi\left(\ppvector{\xi}\right)\mathrm{d}\element + \int_{\partial \element} \left(\ppvector{\flux}\cdot \ppvector{N}\right)\phi\left(\ppvector{\xi}\right)\:\mathrm{d}S - \int_{\element} \ppvector{\flux}\left(\ppvector{U}\right)\cdot\left(\ppvector{\nabla}_{\xi}\:\phi\left(\ppvector{\xi}\right)\right)\mathrm{d}\element= \ppvector{0} \; ,
  \label{eq:weak_formulation}
\end{equation}
where $\ppvector{N}$ is the outward pointing face normal vector. For DGSEM, the solution~$\ppvector{U}$ and the fluxes~$\ppvector{\flux}$ are approximated by polynomial basis functions, which are the tensor product of one-dimensional nodal Lagrangian polynomials $\smash{\ell\left(\xi^k\right)}$, that satisfy the cardinal property $\delta_{ij}$ on a given set of interpolation points $\smash{\left\{\xi^k_j\right\}}$ with $j=0,...,N$ and $k$ indicating the spatial dimension. The solution and the fluxes in three dimensions can therefore be written as:
\begin{align}
  \ppvector{U} \approx \sum_{i,j,k=0}^N\ppvector{\hat{U}}_{ijk}\ell_i\left(\xi^1\right)\ell_j\left(\xi^2\right)\ell_k\left(\xi^3\right) \;,
  \label{eq:polynomial_solution}\\
  \ppvector{\flux} \approx \sum_{i,j,k=0}^N\ppvector{\hat{\flux}}_{ijk}\ell_i\left(\xi^1\right)\ell_j\left(\xi^2\right)\ell_k\left(\xi^3\right) \;.
  \label{eq:polynomial_fluxes}
\end{align}
The test functions are chosen identical to the basis functions and the integrals in \eqref{eq:weak_formulation} are evaluated numerically with collocation of the integration points and interpolation points, for which the Legendre-Gauss and the Legendre-Gauss-Lobatto points are common choices. The choice of shared nodes for both operators leads to a highly efficient numerical scheme with significantly reduced operation counts in 2D and 3D cases.
The gradients of the solution vector for evaluation of the viscous fluxes are obtained with the lifting method by Bassi and Rebay commonly referred to as their "first method"~\cite{bassi1997}. The single elements are only coupled weakly by the fluxes across the elements' faces in the surface integral, which are approximated by a Riemann solver. An appropriate Runge-Kutta method is used to advance the solution in time.\\
Throughout this work all computations were carried out using Legendre-Gauss-Lobatto interpolation points in conjunction with the Split-DG formulation by Pirozolli~\cite{pirozzoli2011} and Roe's approximate Riemann solver \cite{roe1981} with an entropy fix by Harten and Hyman~\cite{harten1983}.
Unless stated otherwise, a fourth-order low-storage Runge-Kutta method~\cite{carpenter1994} is employed for time integration. A complete description of the method, its implementation and parallelization in the code framework FLEXI as well as validation and application examples can be found in~\cite{krais2020flexi}.

\subsection{Non-conforming Meshes}
\label{subsec:non_comforming_meshes}
\begin{figure}[htb]
  \centering
  \begin{subfigure}{0.3\textwidth}
    \raggedright
    \includegraphics[width=0.7\textwidth]{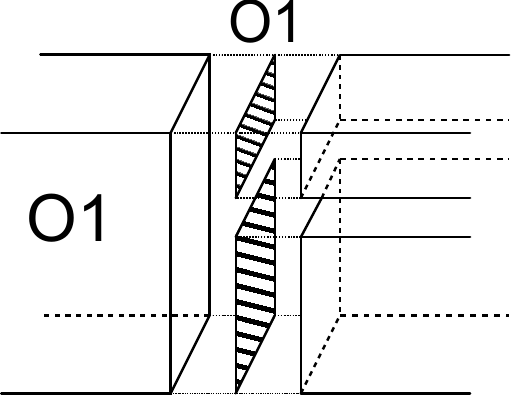}
    \caption{}
    \label{fig:mortar_2D_3D_1}
  \end{subfigure}
  \begin{subfigure}{0.3\textwidth}
    \raggedright
    \includegraphics[width=0.9\textwidth]{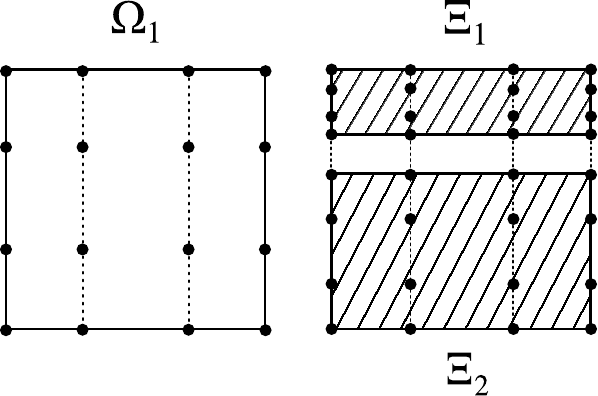}
    \caption{}
    \label{fig:mortar_2D_3D_2}
  \end{subfigure}
  \begin{subfigure}{0.39\textwidth}
    \raggedleft
    \includegraphics[width=0.85\textwidth]{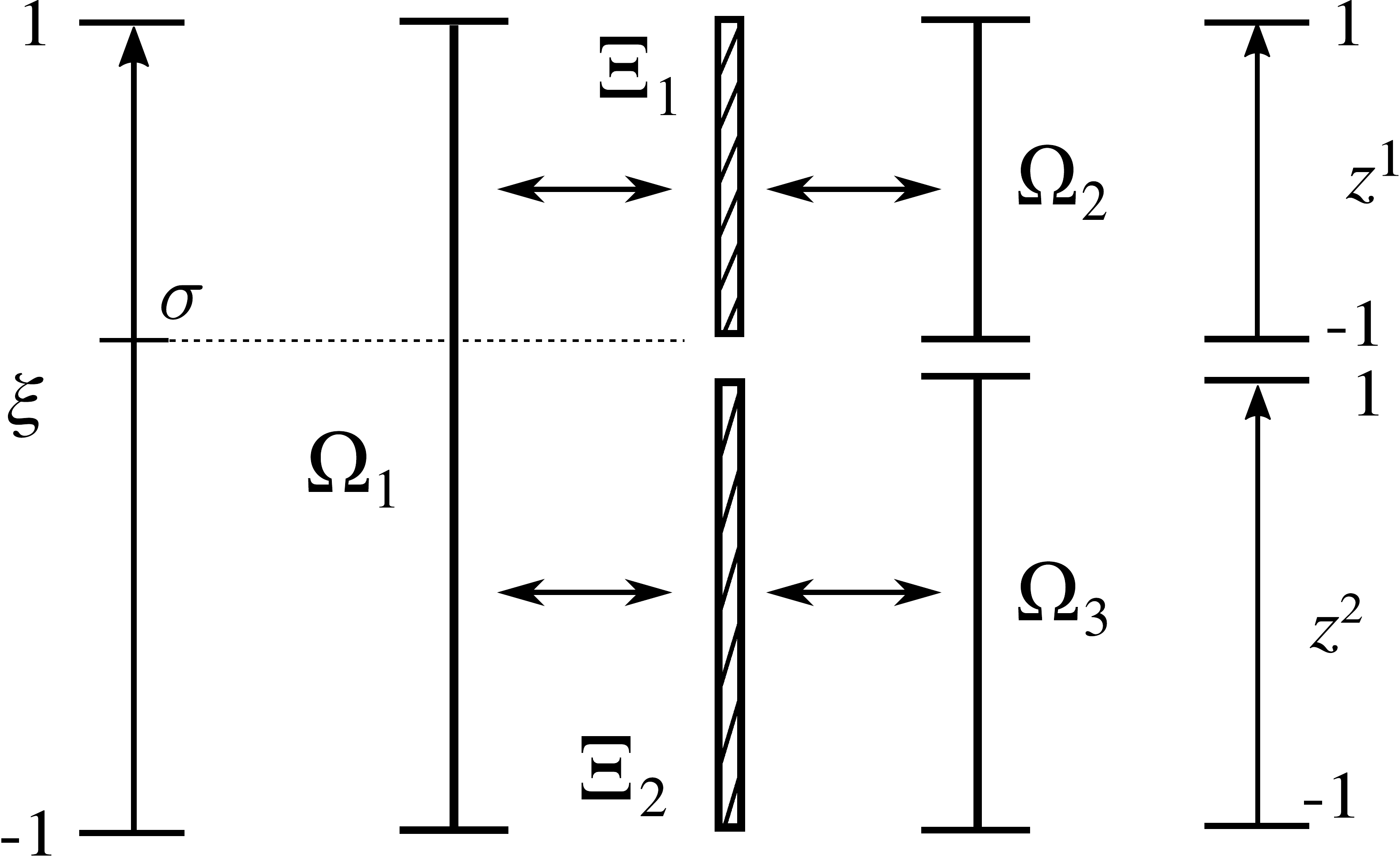}
    \caption{}
    \label{fig:mortar_2D_3D_3}
  \end{subfigure}
  \caption{\textit{Left}: Schematic view of the used mortar method in three dimensions. \textit{Middle}: The distribution of the solution points of the element face~$\Omega_1$ and the mortars~$\Xi_1$,$\Xi_2$ with dotted lines to indicate alignment. \textit{Right:} The resulting quasi-one-dimensional mortar configuration. Additional spacing between the elements is added for clarification and the mortars are shown hatched.}
  \label{fig:mortar_2D_3D}
\end{figure}

The sliding mesh approach, representing the focus of this work, naturally introduces non-conforming element interfaces (sometimes also called hanging nodes) at the sub-domain interface. Even if the initial topology is conforming, the relative movement of adjacent meshes creates a time-dependent interface architecture, in which element neighbors constantly change. In this section, we briefly summarize the static case of such a non-conforming mesh, and extend it to the moving case in~\secref{subsec:sliding_mesh_numerics}.\\
A common approach for the coupling of non-conforming domains in spectral element schemes is the mortar method, originally proposed by Mavriplis in~\cite{mavriplis1989} for the incompressible Navier-Stokes equations and applied to compressible flows by Kopriva in~\cite{kopriva1996,kopriva1998}. It was recently shown by Laughton et al. \cite{laughton2020}, that this mortar approach yields superior accuracy in comparison to interpolation-based methods, especially for underresolved flows.
In the mortar method, the non-conforming interface is subdivided into two-dimensional \textit{mortars} in such a way that each mortar has only one adjacent element face on each side of the interface, as depicted in \figref{fig:mortar_2D_3D}. The sub-domains do not interact directly with each other across the interface. Instead, the solution on each element face at the interface is first projected onto its adjacent mortars. The unique fluxes across the interface are then computed on the mortars and projected back onto the respective element faces on both sides of the interface. A more detailed description of the method implemented in FLEXI can be found in e.g.~\cite{krais2020flexi}.\\
For the configuration depicted in \figref{fig:mortar_2D_3D}, the solution points of the mortars and the element face line up along the dotted lines in~\figref{fig:mortar_2D_3D_2}. This is intended by design - it stems from the fact that we choose the same solution representation on the mortars as for the usual elements. This choice entails that for the chosen tensor product basis, the problem decomposes into individual one-dimensional operations along the dotted lines, as shown in~\figref{fig:mortar_2D_3D_3}. Therefore, only the one-dimensional case along one representative line will be considered in the following.
\subsection*{Projection Domain $\rightarrow$ Mortar}
Analogous to \eqref{eq:polynomial_solution}, the polynomial approximation of the solution on the domain faces $\Omega_k$ is given by
\begin{equation}
  \ppvector{U}^k \approx \sum_{i=0}^N \ppvector{\hat{U}}_{i}^{k} \ell_{i}\left(\xi\right) \;,
  \label{eq:side_solution_ansatz}
\end{equation}
where $\xi$ denotes the one-dimensional coordinate on the element side. To express the solution on the mortars $\Xi_k$, we define a local coordinate $z^k \in \left[-1,1\right]$ such that
\begin{equation}
\xi =
  \begin{cases}
    \begin{aligned}
      \sigma & + \frac{1-\sigma}{2}\left(z^1+1\right) & \;\;\mathrm{for}\;\; & \xi>\sigma\\
      -1     & + \frac{1+\sigma}{2}\left(z^2+1\right) & \;\;\mathrm{for}\;\; & \xi\leq\sigma\\
    \end{aligned}
  \end{cases} ,
\end{equation}
where $\sigma$ denotes the position of the hanging node in reference space as depicted in \figref{fig:mortar_2D_3D_3}.
As shown in~\cite{kopriva1996}, an unweighted $L_2$-projection from the side onto the mortars is sufficient to ensure conservation for straight-edged elements. Using the same basis functions as for the solution $\ppvector{U}$ in \eqref{eq:side_solution_ansatz} on each mortar, the left and right solution of the mortar $\Xi^k$ can be written as
\begin{equation}
  \ppvector{Q}^{\Xi_k,L/R} \approx \sum_{i=0}^N \ppvector{\hat{Q}}_i^{\Xi_k,L/R} \ell_i\left(z^k\right) \;.
  \label{eq:polynomial_solution_mortar}
\end{equation}
Inserting the polynomial representations \eqref{eq:side_solution_ansatz} and \eqref{eq:polynomial_solution_mortar} into the $L_2$-projection of the solution $U^1$ on side $\Omega_1$ onto mortar $\Xi_1$ then reads
\begin{align}
  \int_{-1}^{1}\left(\sum_{i=0}^N\ppvector{\hat{Q}}^{1,L}_i \ell_i\left(z^1\right)-\sum_{i=0}^N\ppvector{\hat{U}}^{1}_i \ell_i\left(\xi\right)\right)\ell_j\left(z^1\right)\:\mathrm{d}z^1 &&& \mathrm{for}\; j=0,...,N \;.
\end{align}
Introducing the definitions
\begin{align}
  M_{ij} &:= \int_{-1}^1 \ell_i\left(z^1\right)\ell_j\left(z^1\right)\mathrm{d}z^1 && \mathrm{for}\; i,j=0,...,N \;,
\label{eq:proj_M}\\
  S_{ij}^{\Omega_1\rightarrow\Xi_1} &:= \int_{-1}^1 \ell_i\left(\xi(z^1)\right)\ell_j\left(z^1\right)\mathrm{d}z^1 && \mathrm{for}\; i,j=0,...,N \;,
\label{eq:proj_S}
\end{align}
finally leads to the projection operation as
\begin{equation}
  \ppvector{\hat{Q}}^{1,L} = \ppmatrix{M}^{-1}\ppmatrix{S}^{\Omega_1\rightarrow\Xi_1} \ppmatrix{\hat{U}}^1 := \ppmatrix{P}^{\Omega_1\rightarrow\Xi_1} \ppmatrix{\hat{U}}^1 \;,
  \label{eq:projection_domain_mortar}
\end{equation}
where $\ppmatrix{P}^{\Omega_1\rightarrow\Xi_1}$ is defined as the projection matrix from face~$\Omega_1$ to mortar~$\Xi_1$. The projection from face~$\Omega_1$ onto mortar~$\Xi_2$ can be obtained accordingly. In addition, it seems worth noting that the mass matrix~$\ppmatrix{M}$ depends only on the used basis functions and is therefore independent of the interface configuration.
Furthermore, the projection from the element faces onto the mortars is formally exact, provided that the same polynomial degree $N$ for the solution on mortars and element faces is employed.
Since the lifting routine employs a DGSEM discretization for the gradients, these can be obtained by the same procedure as is described here. The resulting gradients on the element faces can be projected onto the mortars analogously using \eqref{eq:projection_domain_mortar}.

\subsection*{Projection Mortar $\rightarrow$ Domain}
Once the left and the right solution on the mortars $\Xi_1$ and $\Xi_2$ are obtained, the left and right fluxes $\ppvector{\flux}^{L/R,\Xi_{1/2}}$ can be evaluated. The resulting fluxes are projected back onto the domain face $\Omega_1$ using an $L_2$-projection in the form of
\begin{equation}
  \int_{-1}^{\sigma} \left(\ppvector{\flux}^{\Omega_1}\left(\xi\right)-\ppvector{\flux}^{\Xi_1,L}\left(z^1\right)\right)\ell_j\left(\xi\right)\mathrm{d}\xi + \int_{\sigma}^{1} \left(\ppvector{\flux}^{\Omega_1}\left(\xi\right)-\ppvector{\flux}^{\Xi_2,L}\left(z^2\right)\right)\ell_j\left(\xi\right)\mathrm{d}\xi = 0 \;\;\mathrm{for}\;j=0,...,N \;.
\end{equation}
By inserting the polynomial representation \eqref{eq:polynomial_fluxes} of the fluxes and reordering we obtain
\begin{equation}
  \ppvector{\hat{\flux}}^{\Omega} = \ppmatrix{P}^{\Xi_1\rightarrow\Omega}\:\ppvector{\hat{\flux}}^{\Xi_1,L} + \ppmatrix{P}^{\Xi_2\rightarrow\Omega}\:\ppvector{\hat{\flux}}^{\Xi_2,L} := \left(\tfrac{1-\sigma}{2}\right) \ppmatrix{M}^{-1}\ppmatrix{S}^{\Xi_1\rightarrow\Omega}\:\ppvector{\hat{\flux}}^{\Xi_1,L} + \left(\tfrac{1+\sigma}{2}\right)\ppmatrix{M}^{-1}\ppmatrix{S}^{\Xi_2\rightarrow\Omega}\:\ppvector{\hat{\flux}}^{\Xi_2,L} \;,
  \label{eq:projection_mortar_domain}
\end{equation}
where $\ppmatrix{M}$ is identical to the matrix defined in \eqref{eq:proj_M} and the matrices $\ppmatrix{S}^{\Xi_1\rightarrow\Omega}$ and $\ppmatrix{S}^{\Xi_2\rightarrow\Omega}$ are the transposes of $\ppmatrix{S}^{\Omega\rightarrow\Xi_1}$ and $\ppmatrix{S}^{\Omega\rightarrow\Xi_2}$ in \eqref{eq:proj_S}, respectively.

\subsection{Sliding Mesh Interface}\label{subsec:sliding_mesh_numerics}
We briefly lay out the sliding mesh idea as proposed and described in detail in \cite{zhang2015}. At a sliding mesh interface, two sub-domains perform a sliding relative motion along the interface. For simplicity, it can be assumed that one sub-domain is static while the other is moving (although in principle, both sub-domains can be moving). We first consider two-dimensional cases, i.e. where the interface is a 1D object, and we restrict ourselves to periodic interfaces. For planar interfaces, periodicity can be ensured with periodic boundary conditions. The alternative are circular interfaces, where one sub-domain is inside the circle, and the other outside, and the relative motion is a rotation.  
\begin{figure}[htb!]
  \centering
  \includegraphics[width=0.2\textwidth]{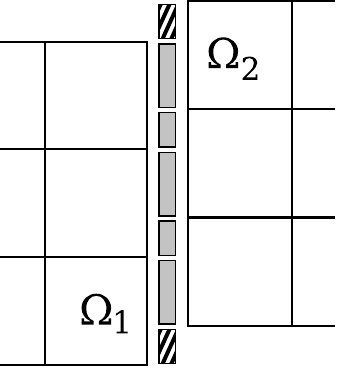}
  \caption{Schematic of the planar sliding mesh interface. The gap between the domains allows to show the mortar configuration at the interface (gray). The striped mortars are identical for periodic boundaries at the interface.}
  \label{fig:SlidingMesh_Interface}
\end{figure}
Starting from an initial configuration where element faces are conforming and equi-spaced along the interface, the relative motion leads to a non-conforming pattern, as shown in~\figref{fig:SlidingMesh_Interface}. We note that our algorithm can also work with non-equispaced interface elements, but for all the applications envisioned by us, there is no reason to suggest that such a mesh topology is necessary. We will thus stick to the described spacing for the rest of this work.
 Extending the idea for the static case from~\secref{subsec:non_comforming_meshes} to the moving case follows these steps: 
\begin{enumerate}
  \item Introduce a dynamic definition of the interface configuration.
  \item Introduce two-sided mortars between the non-conforming sub-domains as shown in~\figref{fig:SlidingMesh_Interface}.
  \item Interpolate the solution from the faces of both sub-domains onto these mortars.
  \item Solve the Riemann problem on the mortars to obtain numerical fluxes.
  \item Project the numerical flux of the mortars back onto the faces of both sub-domains.
\end{enumerate}
Due to the relative motion, the interface is now necessarily dynamic, i.e. once the definition has been updated to reflect that situation in item 1, the rest of the algorithm can follow the static procedure for this instance in time, and compute the appropriate projection and interpolation operators. The dynamic definition of the interface has to account for two aspects: Firstly, the size of the overlap regions of elements changes, i.e. the mortar definition has to be adjusted accordingly. Secondly, the neighboring information across the interface is now also dynamic and changes during the computation.  The restriction to equidistant spacing at the interface has two consequences: Each element face is always represented by two mortars (with the exception of the singular moments of conforming sub-domains, which are in practice handled by one mortar of the size of the element faces, and the other of size $0$). Moreover, the position of the hanging nodes $\sigma$ in reference space is the same for all faces, such that the same interpolation and projection matrices can be used for all faces along the interface.
\begin{figure}
  \centering
  \begin{subfigure}[b]{0.2\textwidth}
    \centering
    \includegraphics[width=0.8\textwidth]{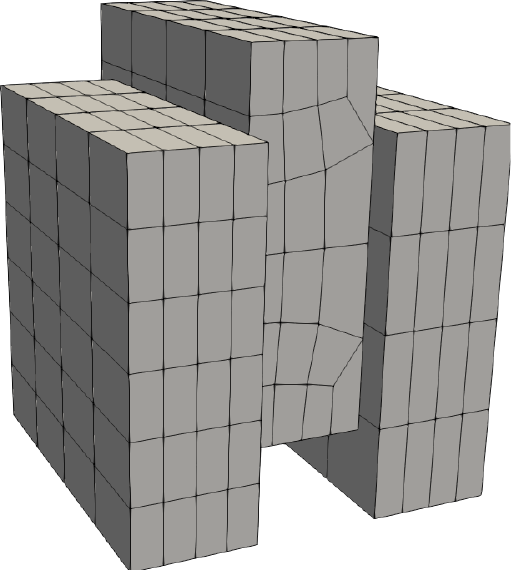}
  \end{subfigure}
  \hspace{0.05\textwidth}
  \begin{subfigure}[b]{0.2\textwidth}
    \centering
    \includegraphics[width=0.7\textwidth]{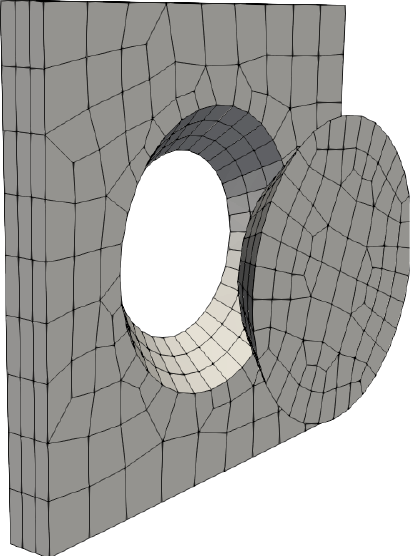}
  \end{subfigure}
  \hspace{0.05\textwidth}
  \begin{subfigure}[b]{0.2\textwidth}
    \centering
    \includegraphics[width=0.75\textwidth]{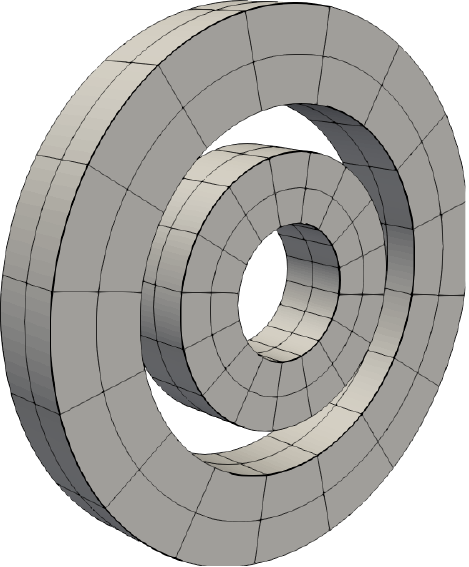}
  \end{subfigure}
  \hspace{0.05\textwidth}
  \begin{subfigure}[b]{0.2\textwidth}
    \centering
    \includegraphics[width=0.95\textwidth]{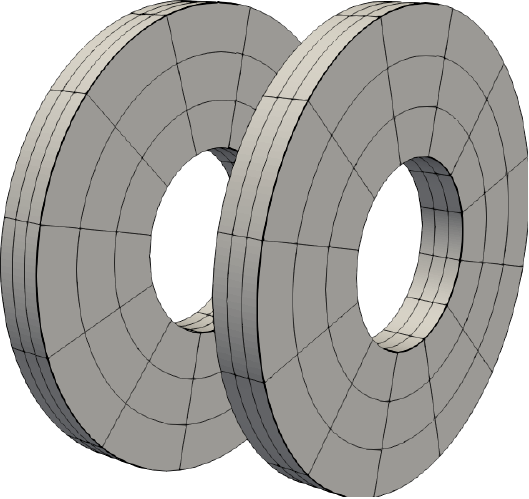}
  \end{subfigure}
  \caption{Possible mesh geometries in three spatial dimensions for the described sliding mesh method. The moving sub-domains are scaled or shifted to reveal the structured mesh at the interfaces. From left to right the interface geometries allow for translational movement, rotation with a conical interface, rotation with a cyclindrical radial interface and rotation with a plane annular axial interface.}
  \label{fig:sm_meshes_overview}
\end{figure}\\
For three-dimensional domains, the interfaces become two-dimensional objects. A coordinate perpendicular to the interface movement can be introduced (in principle, the relative movement is not confined to one coordinate, but in the present work, we restrict ourselves to this case). Perpendicular to the movement, several element layers can be introduced, but the face mesh at the interface has to be structured. For circular and planar interfaces, three-dimensional equivalents are shown in \figref{fig:sm_meshes_overview} along with an axial interface of two annular sub-domains with relative rotation.

\section{Parallel Sliding Mesh Implementation}
\label{sec:implementation}
With the mathematical operators for the sliding mesh interface in place, we now present a strategy for an efficient implementation in our in-house code FLEXI and possibly other element-based high-order schemes. It is designed to minimize the thread-level computational overhead, but more importantly to keep communication as efficient as possible. To this end, global communication is avoided altogether and local communication is kept to a minimum by passing only data and no metadata like indices or identifiers.

In \secref{sec:implementation_flexi}, a very brief introduction to the MPI-based parallelization strategy of our baseline code is given. The basic approach to the sliding mesh implementation is given in \secref{sec:implementation_basics}. The most challenging aspect of the parallelization is to generate information about size and sorting of a set of data for each passed message in a setting where the communication partners are dynamically changing. To facilitate the description of our approach to this dynamic configuration, some index definitions are introduced in~\secref{sec:implementation_definitions}. On this basis, the data sorting and index mapping is described in \secref{sec:implementation_sorting}. An illustrative example for the index mapping is given in \ref{sec:example}.

\subsection{Prerequisites: FLEXI Parallelization}
\label{sec:implementation_flexi}
In order to formulate requirements to the sliding mesh implementation, some principles of the underlying FLEXI code are first briefly laid out. FLEXI uses a pure distributed memory (MPI) parallelization.
In the mesh building process using our in-house High-Order Pre-Processor (HOPR)~\cite{hindenlang2015}, elements are sorted along a space filling curve~\cite{krais2020flexi}. During mesh decomposition in FLEXI, one or several complete DG elements are assigned to each process following the sorting along the space filling curve. This allows to obtain compact sub-domains for each process. No element is split between two processors.
At each interface between two elements, one of the elements is defined as primary and the other as replica with respect to that interface. We note that the connection between elements in a DG scheme is achieved by the numerical flux function akin to a finite volume scheme. If the two elements adjacent to an interface are handled by two different processes, two communication steps are to be carried out during each computation of a DG operator to compute a common interface flux: First, the solution $U$ on the boundary is passed from the replica to the primary element. The Riemann flux is calculated on the primary element and the result is passed back to the replica element (two more analogous communication steps are required for the lifting procedure for the computation of viscous terms). The solution $U$ and the fluxes $F$ at the boundary are stored in separate arrays for primary and replica (yielding four arrays $U_\text{primary}$, $U_\text{replica}$, $F_\text{primary}$ and $F_\text{replica}$). In these arrays, the values are sorted according to an element interface index $\flexiSideID$, which is assigned to each element interface during the initialization phase of the simulation. $\flexiSideID$ is process-local.

For the interfaces forming a process interface, interface indices $\flexiSideID$ are ordered nestedly by two criteria:
\begin{itemize}
  \item The outer sorting is by the rank of the neighboring process. This ensures the data sent to one process to be contiguous in memory.
  \item The inner sorting (of all element interfaces belonging to the same neighboring process) is by a criterion unique for all processes (in our case, a global face index stored in the mesh file for each element interface). This inner, globally unique sorting ensures that the order of the data communicated between two processes is consistent.
\end{itemize}
A similar strategy will employed for the sliding mesh interface.

In order to achieve high compute throughputs without having to wait for communications to finish, latency hiding is employed in FLEXI: The operations necessary in preparation of a communication step are always carried out first at the earliest possible instance. Also, the communication is initiated as early as possible in a non-blocking manner. The communication window is then filled with local arithmetic operations to give the message passing as much time as possible to complete.

\subsection{Sliding Mesh: Implementation Basics}
\label{sec:implementation_basics}

For each sliding mesh interface, there is an adjacent static and a moving mesh sub-domain. The MPI domain decomposition occurs in two steps: First, each process is assigned to either the sliding or the moving domain, so that no process handles elements on both sides of a sliding mesh interface. This choice eases implementation, but also increases efficiency, as a processor sub-domain across a sliding mesh interface would get torn apart and lose its compact shape due to the movement. Within each of these sub-domains, elements are already sorted along a space filling curve during mesh generation and the elements handled by each process are assigned accordingly.

The elements belonging to the static domain are defined to be the primary elements for the sliding mesh interface. The solution $U$ is first interpolated from the element faces to the mortars on both sides of the interface. The additional mortar arrays $U_\text{primary}^\text{sm}$ and $U_\text{replica}^\text{sm}$ exist for this purpose, alongside the additional arrays $F_\text{primary}^\text{sm}$ and $F_\text{replica}^\text{sm}$. The communication procedure is similar to the one for conforming faces: The solution $U_\text{replica}^\text{sm}$ is passed from the moving (replica) to the static (primary) domain, where the Riemann flux is evaluated. The flux $F_\text{replica}^\text{sm}$ is passed back to the moving (replica) process and on both sides of the interface, the flux is projected onto the DG basis for the respective element face. Communication hiding is employed in the same manner as for the standard conforming element interfaces.

\begin{figure}
  \centering
  \begin{minipage}{.45\textwidth}
    \centering
    \includegraphics[width=0.80\linewidth]{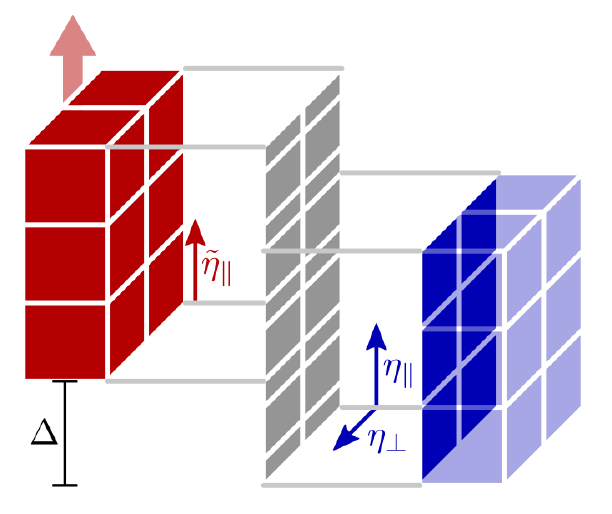}
  \end{minipage}\hspace*{0.05\textwidth}
  \begin{minipage}{.45\textwidth}
    \centering
    \includegraphics[width=0.5\linewidth]{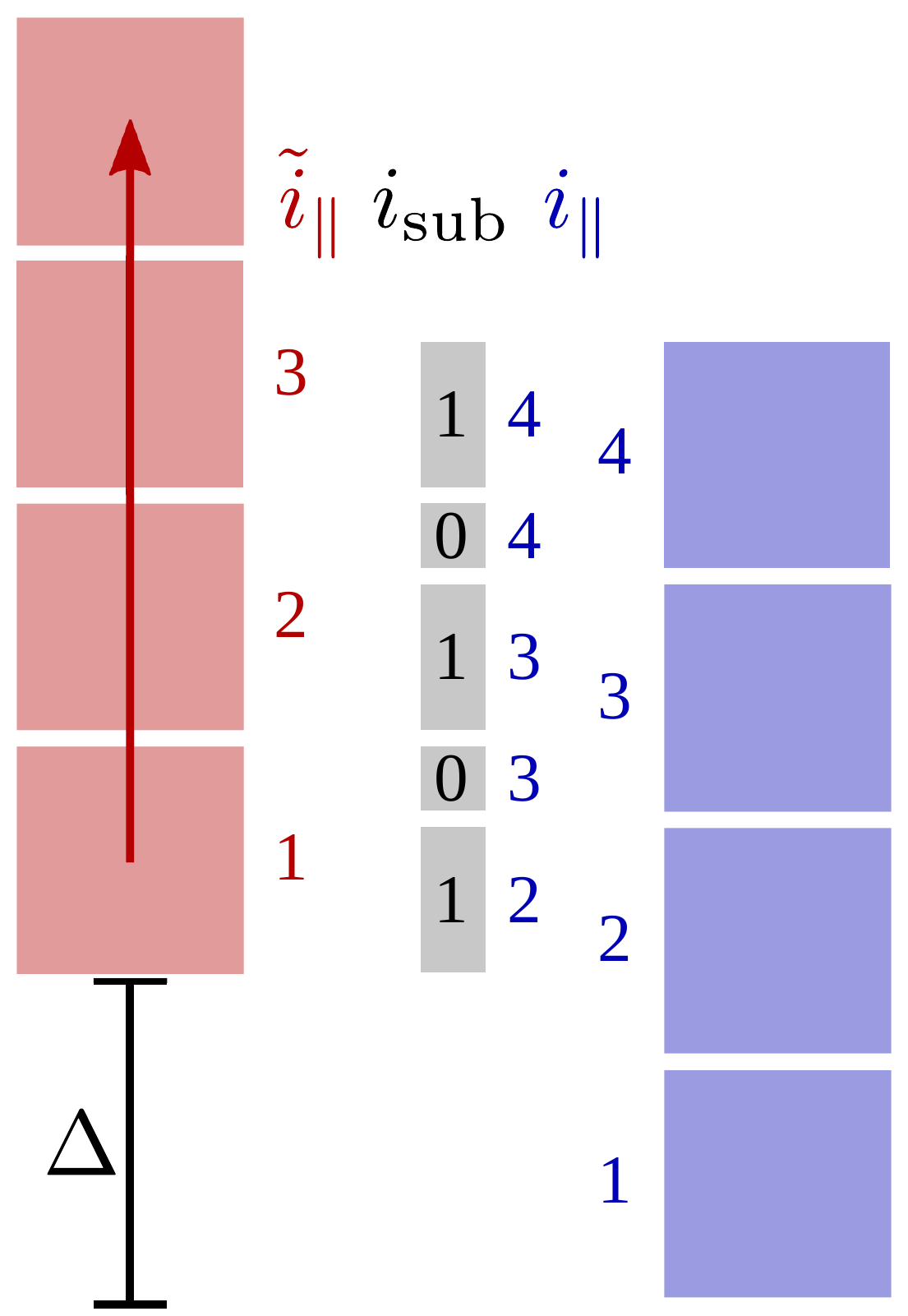}
  \end{minipage} \newline
  \begin{minipage}[t]{.45\textwidth}
    \captionof{figure}{Mortar structure and coordinate definitions for a planar interface with periodic boundaries. The left (red) sub-domain is moving vertically, resulting in a displacement $\disp$. The right (blue) sub-domain is static. As shown, $\mcopa = \scopa-\disp$ for the parallel coordinates.}
    \label{fig:coordinates}
  \end{minipage}\hspace*{0.05\textwidth}
  \begin{minipage}[t]{.45\textwidth}
    \captionof{figure}{Index definition example for a planar interface. Left (red) sub-domain is moving, right (blue) is static. Plane view: the normal coordinate and the according index are omitted for clarity. Note that the mortars use the index $\sipa$ of the static sub-domain.}
    \label{fig:indices}
  \end{minipage} \\
\end{figure}

\subsection{Parallelization: Index Definitions}
\label{sec:implementation_definitions}

In the following, some index definitions are introduced to ease the description of index mapping and sorting. Particularly, it will be necessary to uniquely address each mortar, each static and each moving face by a set of indices.
The following definitions are also illustrated in Figures \ref{fig:coordinates} and \ref{fig:indices}.
Variables on the static side (and in a static un-displaced frame of reference) are noted without an accent, while variables on the moving side (and in a frame of reference displaced with the moving sub-domain) are marked with a tilde $\mov{\cdot}$. Let us first consider the static and un-displaced frame of reference: We define the coordinates $\scopa$ and $\scope$
at the sliding mesh interface, where the subscripts $\parallel$ and $\perp$ indicate their direction relative to the mesh movement (cf. \figref{fig:coordinates}).
The faces at the interface are placed in a structured grid, so two indices $\sipa$ and $\sipe$ can be assigned to each static element face at the interface, numbering them along $\scopa$ and $\scope$.

On the moving side, a parallel coordinate $\mcopa$ displaced with the moving sub-domain and an according index $\mipa$ are introduced, such that $\mcopa = \scopa-\disp$, where $\disp$ is the displacement of the moving domain. This is illustrated in \figref{fig:coordinates}. Since there is no displacement in the perpendicular direction, the coordinate $\scope$ and the index $\sipe$ can be used for the moving domain, too, and no additional corresponding variables for the moving side have to be introduced. Faces on the moving side are uniquely defined by the two indices $\mipa$ and $\sipe$. The displacement $\disp$ can be expressed in terms of the number of surpassed faces $n_\disp$ and the fraction of the currently surpassed face $s_\disp$, i.e.
\begin{align}
  \disp = (n_\disp + s_\disp)\,\lpa, \quad n_\disp\in\Z,
\end{align}
where $\lpa$ is the face length along the direction of sub-domain movement.

The mortars inherit the index from the{\it~static} domain $\sipa$ (as well as $\sipe$). In order to uniquely define each mortar, a third index $\im\in\{0,1\}$ is introduced to distinguish between the two mortars adjacent to an element face. It is defined as
\begin{align}
   \im =
  \begin{cases}
    0 & \text{for}\quad \scopa-\sipa\lpa<s_{\disp}\lpa,\\
    1 & \text{else.}
  \end{cases}\label{eq:subim}
\end{align}
Following this definition, for each face on the static side, the mortar with the smaller $\scopa$ (i.e. the "lower" mortar in \figref{fig:indices}) has index $\im=0$ and the "upper" one has $\im=1$, while the order is inverse on the moving side. The indices $\mipa$, $\sipa$ and $\im$ of a mortar and the adjacent faces are finally linked via the relation
\begin{align}
  \mipa = \sipa - n_\disp + \im - 1, \label{eq:indices}
\end{align}
which can be exemplarily verified in \figref{fig:indices} for $n_\disp = 1$. The normal index $\sipe$ is of course the same for a mortar and its adjacent element faces.

\subsection{Parallelization: Mortar Sorting and Index Mapping}
\label{sec:implementation_sorting}

Communication in the presence of changing mesh topology and changing communication partners poses unique challenges. In order to avoid additional communication, several requirements have to be met:
\begin{enumerate}
  \item The communication partners' ranks as well as the size of the communicated data sets need to be known a priori by both sides.
  \item The data communicated from one process to another should be contiguous in memory on both processes.
  \item The data communicated from one process to another has to be sorted by universal criteria, such that the receiving process knows beforehand how the data is sorted.
\end{enumerate}

We start by addressing the first requirement. To this end, the ranks handling the elements belonging to each static face $\stat{\face}_{\sipa,\sipe}$ and each moving face $\mov{\face}_{\mipa,\sipe}$ are communicated globally during the initialization phase prior to the actual simulation and are stored as two mapping arrays $\stat{\rank}(\sipa,\sipe)$ and $\mov{\rank}(\mipa,\sipe)$. This is the only global communication procedure in the proposed implementation. Subsequently, all dynamic information regarding the configuration of the interface and the partners in a communication step across the interface can be deduced without the need for further message passing.

We now have all necessary ingredients in place to meet the above three criteria:
\begin{enumerate}
  \item For each mortar, the ranks handling both adjacent faces are known via $\stat{\rank}$ and $\mov{\rank}$ (we use \eqref{eq:indices} to translate between $\mipa$ and $\sipa$).
  \item Sorting the mortars on each process by its communication partner yields contiguous data chunks to be passed.
  \item The index triple $(\sipa,\sipe,\im)$ for each mortar is a globally unique criterion for the inner sorting of the communicated data sets.
\end{enumerate}

The sorting procedure now works as follows: An index array $\sortedArray$ is set up, which contains a tuple of five entries for each mortar adjacent to the faces of the own rank. On each process of the{\it~static} domain, these entries are: the FLEXI face indices $\flexiSideID$, the ranks of the opposing{\it~moving} domain ranks $\mrank$, the movement-parallel indices $\sipa$, the normal indices $\sipe$, and the mortar index $\im$. The only difference on the{\it~moving} domain is that here, the ranks of the{\it~static} processes $\srank$ are stored instead of $\mrank$. These index arrays are then nestedly sorted by the four indices $\srank$ (or $\mrank$, respectively), $\sipa$, $\sipe$ and $\im$ (from outer to inner in that order) using a quicksort algorithm. The FLEXI face index $\flexiSideID$ is passively sorted by the other variables. The resulting order determines the sorting of mortar data on both the static and the moving side.
The order of the entries in $\sortedArray$ defines a process-local index $\iMortar$, which determines the data sorting in the arrays
$U_\text{primary}^\text{sm}$, $U_\text{replica}^\text{sm}$, $F_\text{primary}^\text{sm}$ and $F_\text{replica}^\text{sm}$.

An index array $\mapping$ is set up, where $\iMortar$ is given for each $\flexiSideID$ and $\im$. It is used to store data ordered during the interpolation and the projection steps of the mortar procedure.

The sorting and mapping process is illustrated with the help of an example in \ref{sec:example}.

Updates of the index arrays $\sortedArray$ and $\mapping$ are only carried out whenever the communication structure changes, i.e. whenever the moving sub-domain displacement surpasses a full face length. This is, in fact, the only necessary procedure to account for the changed mesh topology. At all other time stages, $\mapping$ remains the same and only $s_\disp$ changes, so only the interpolation and projection operators have to be updated.

For extension to multiple sliding mesh interfaces, this strategy can be performed individually for each interface. The mortar sorting can then be kept unique by introducing a new (outermost) sorting index to distinguish between interfaces.

\section{Results}
\label{sec:results}
In this section, the high-order accuracy of the implemented sliding mesh method is verified by convergence tests for a curved interface in \secref{subsec:convergence_curved} and a straight interface in \secref{subsec:convergence_planar}. We then investigate the parallel performance of the novel method and compare it against the static baseline scheme in~\secref{subsec:scaling}, before the method is applied to a large scale LES test case in \secref{subsec:turbine}.\\

\subsection{Isentropic Vortex}
\label{subsec:convergence_curved}
To verify that the implemented method is indeed high-order accurate for DGSEM and especially for curved interfaces which are approximated by high-order polynomials, we follow Zhang and Liang \cite{zhang2015} and investigate the order of accuracy of the method using the transport of an isentropic Euler vortex.
For this two-dimensional test case, an isentropic vortex is superimposed on a constant freestream. For details on the exact solution and the notation we refer the reader to~\cite{zhang2015}. The vortex parameters are chosen as $\epsilon=1$, $r_c=1$, which can be interpreted as vortex intensity and vortex size, respectively. At the beginning $t=0$, the vortex is located in the center of the domain. The freestream is initialized with $\rho_{\infty}=1$, $v_{\infty}=1$, $\theta=\mathrm{arctan\left(\tfrac{1}{2}\right)}$, $Ma_{\infty}=0.3$ as the freestream density, velocity magnitude, flow angle and the freestream Mach number, respectively. The ratio of specific heats is set to $\gamma=1.4$ and the freestream pressure is set consistent with the Mach number via the ideal gas relation.
\begin{figure}[htb]
  \begin{subfigure}{0.49\textwidth}
    \centering
    \includegraphics[height=0.6\textwidth]{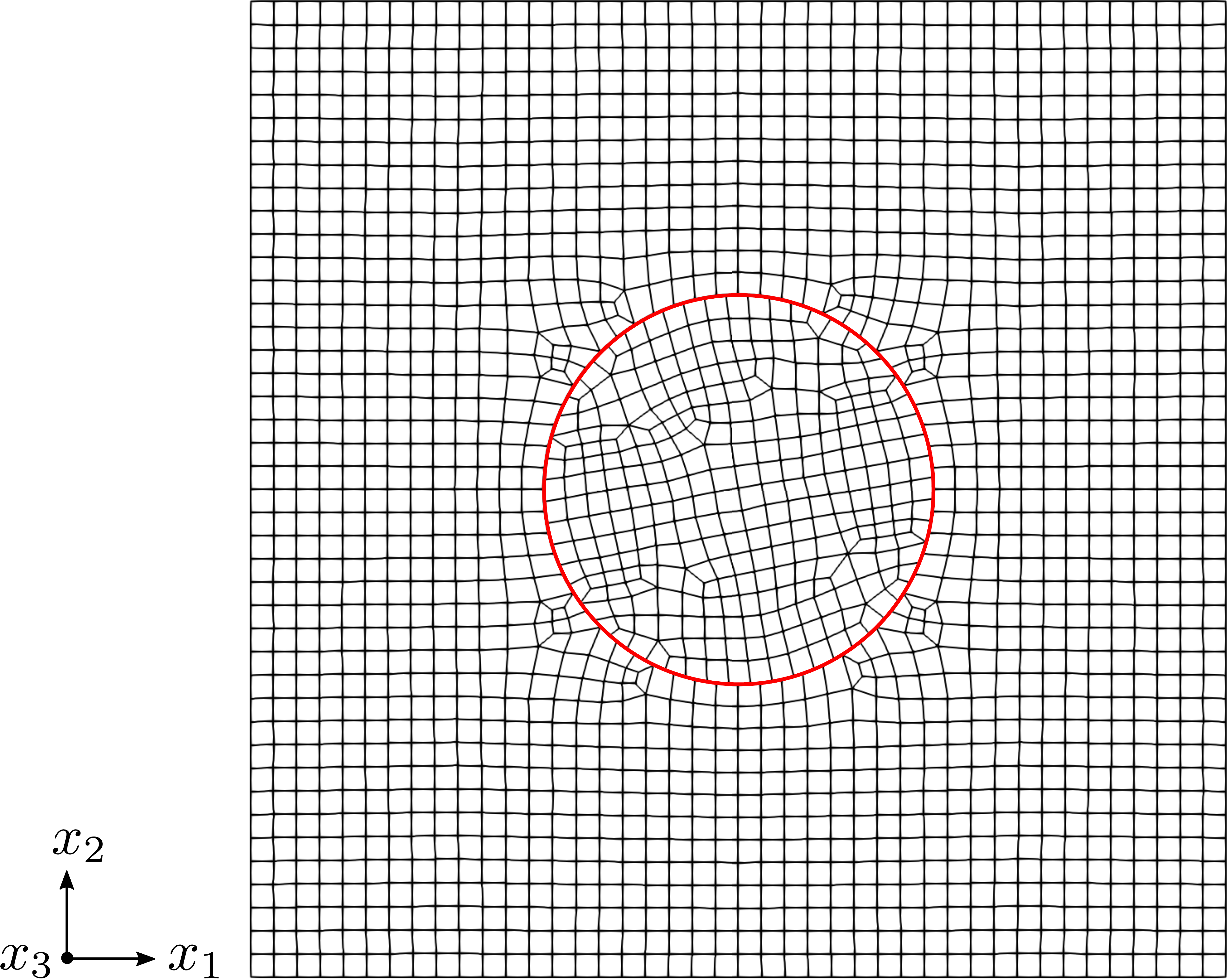}
  \end{subfigure}
  \begin{subfigure}{0.49\textwidth}
    \centering
    \includegraphics[height=0.6\textwidth]{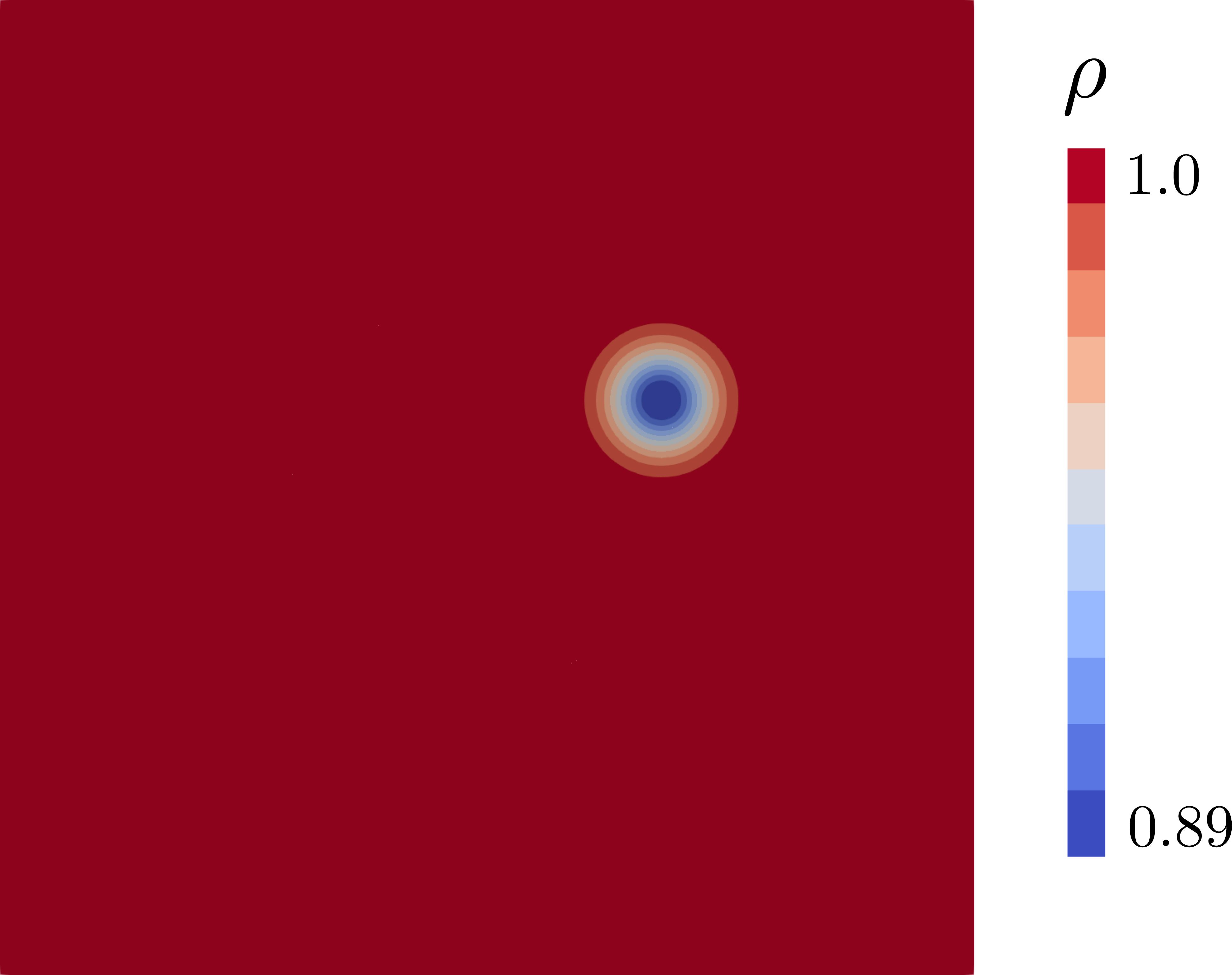}
  \end{subfigure}
  \caption{\textit{Left:} The coarsest mesh with $1854$ elements at $t=4.0$, the sliding mesh interface is highlighted red. \textit{Right:} Exact solution of the isentropic vortex for the density $\rho$ at $t=4.0$.}
  \label{fig:vortex_convergence}
\end{figure}\\
Three different unstructured meshes with the number of elements ranging between $1854$ and $15003$ elements are used, with the coarsest one depicted in \figref{fig:vortex_convergence} together with the exact solution. The meshes are unstructured and quadratic with a side length of $L=20$ and periodic boundary conditions. The inner sub-domain rotates with an angular velocity of $\omega=0.1$. All errors in \tabref{tab:Conv_Vortex} are reported at $t=4.0$ when the center of the vortex reaches the sliding mesh interface. For all cases, the explicit time step is reduced artificially to highlight the behaviour of the spatial discretization error.
The method indeed shows the expected convergence behavior for all investigated orders.
\begin{table}[htb!]
  \centering
  \setlength{\tabcolsep}{10pt}
  \begin{tabularx}{\textwidth}{rXlclclclc}
    \toprule
               && N=2         &       & N=3         &       & N=4         &       & N=5         &       \\
    \#Elem     && $L_2$-Error & Order & $L_2$-Error & Order & $L_2$-Error & Order & $L_2$-Error & Order \\
    \midrule
    $ 1854$    && 8.58e-05    & ---   & 8.60e-06    & ---   & 1.09e-06    & ---   & 1.51e-07    & ---   \\
    $ 7094$    && 1.52e-05    & 2.58  & 8.40e-07    & 3.47  & 4.94e-08    & 4.61  & 3.13e-09    & 5.78  \\
    $15003$    && 4.35e-06    & 3.34  & 1.65e-07    & 4.35  & 7.71e-09    & 4.96  & 3.85e-10    & 5.60  \\
    \bottomrule
  \end{tabularx}
  \caption{$L_2$-Errors of density $\rho$ for the isentropic vortex at $t=4.0$ and the resultant orders of accuracy for several polynomial degrees $N$.}
  \label{tab:Conv_Vortex}
\end{table}

\subsection{Manufactured Solution}
\label{subsec:convergence_planar}
\begin{figure}[htb]
  \begin{subfigure}{0.49\textwidth}
    \centering
    \includegraphics[height=0.2\textheight]{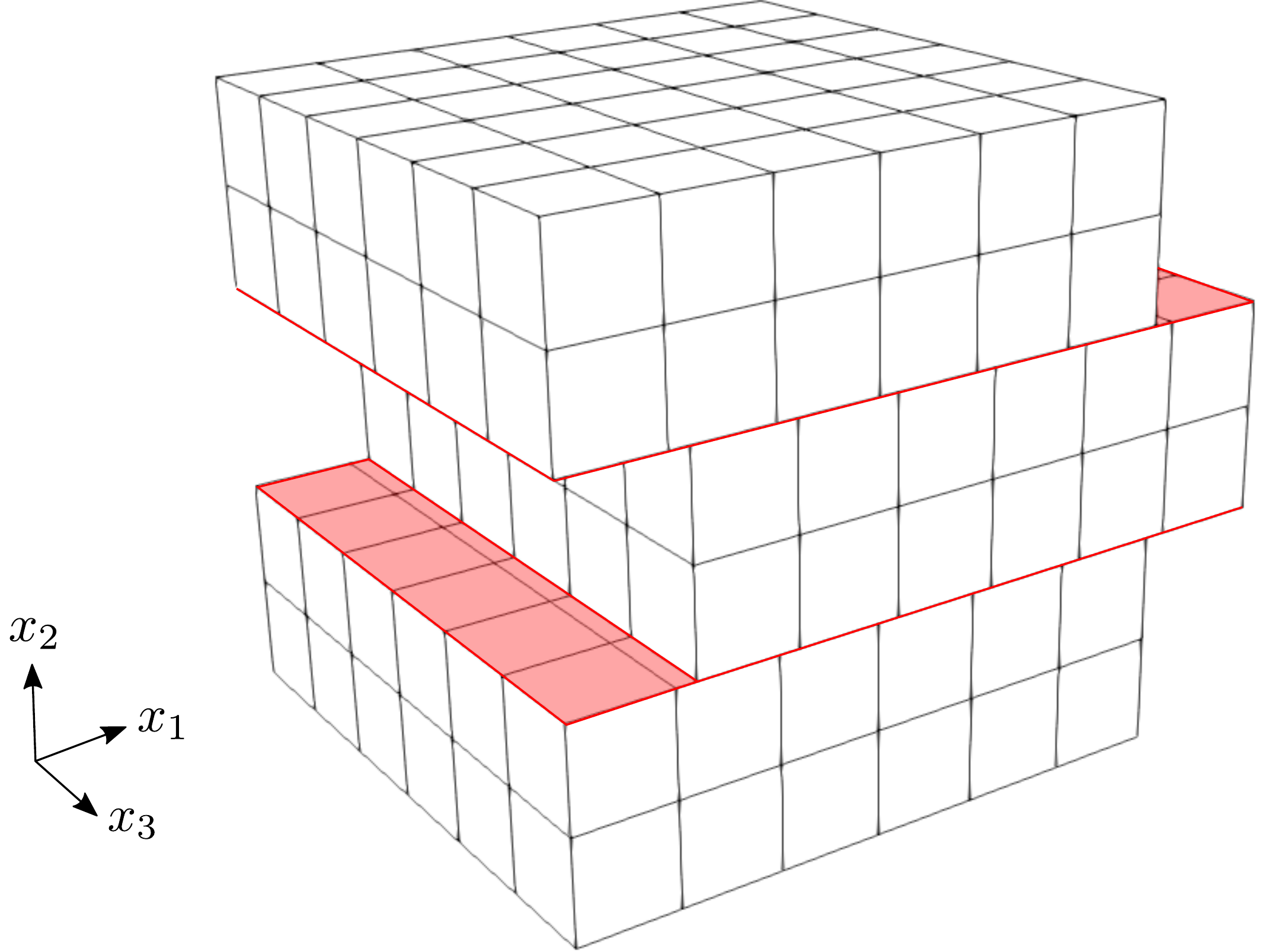}
  \end{subfigure}
  \begin{subfigure}{0.49\textwidth}
    \centering
    \includegraphics[height=0.2\textheight]{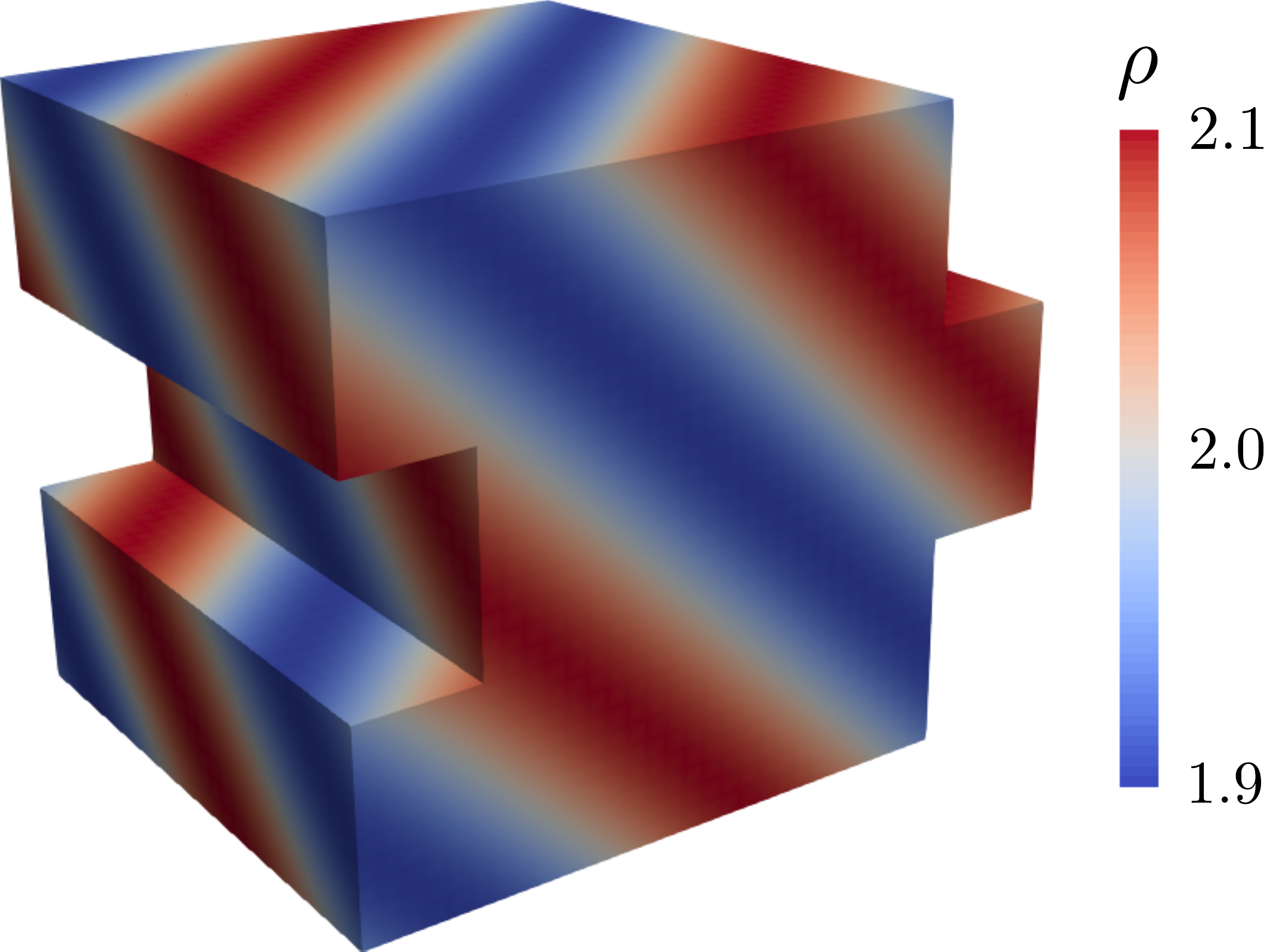}
  \end{subfigure}
  \caption{\textit{Left:} Computational mesh with $6^3$ elements at $t=0.3$, the sliding mesh interfaces are highlighted in red. \textit{Right:} Exact solution of the manufactured solution for the density $\rho$ at $t=0.3$.}
  \label{fig:sine_convergence}
\end{figure}

\renewcommand{\ppvector}[1]{\ensuremath{\vec{#1}}}
To verify the high-order accuracy of the implemented method for fully three-dimensional flows, a smooth manufactured solution from \cite{gassner2009} is investigated. The corresponding parameters are set to $\omega=1$, $\alpha=0.1$, $R=287.058$, $\mu=0.001$. This manufactured solution describes an oblique sine wave advected with a constant speed, as shown in \figref{fig:sine_convergence}.
The computational domain is set to $\ppvector{x}\in\left[0,2\right]^3$ with a Cartesian mesh and periodic boundary conditions, while the central sub-domain (i.e. the sub-domain $x_2\in\left[\tfrac{2}{3},\tfrac{4}{3}\right]$) moves with a velocity of $\ppvector{v}^g=\left[1,0,0\right]^T$, as depicted in \figref{fig:sine_convergence}. Starting from a cube with $3^3$ elements, the number of elements in each spatial direction is doubled in every refinement step.
\\
The time step is chosen small enough to inhibit any influence of the time integration scheme on the observed errors.
The $L_2$-errors for the density $\rho$ at $t=1.0$ and the resultant orders of accuracy are reported in \tabref{tab:Conv_Sine}.
As before, the sliding mesh method retains the high-order accuracy of the DGSEM scheme as expected.
\begin{table}[htb!]
  \centering
  \setlength{\tabcolsep}{10pt}
  \begin{tabularx}{\textwidth}{rXlclclclc}
    \toprule
              && N=2         &       & N=3         &       & N=4         &       & N=5         &       \\
    \#Elem    && $L_2$-Error & Order & $L_2$-Error & Order & $L_2$-Error & Order & $L_2$-Error & Order \\
    \midrule
    $ 3^3$    && 4.21e-02    &  --   & 5.08e-03    &  --   & 3.16e-04    &  --   & 2.98e-05    &  --   \\
    $ 6^3$    && 3.82e-03    & 3.46  & 1.60e-04    & 4.99  & 9.80e-06    & 5.01  & 6.86e-07    & 5.44  \\
    $12^3$    && 4.93e-04    & 2.95  & 1.02e-05    & 3.97  & 3.51e-07    & 4.81  & 1.07e-08    & 6.00  \\
    $24^3$    && 6.93e-05    & 2.83  & 6.79e-07    & 3.91  & 1.13e-08    & 4.95  & 1.64e-10    & 6.03  \\
    $48^3$    && 8.47e-06    & 3.03  & 4.44e-08    & 3.93  & 3.05e-10    & 5.21  & 2.36e-12    & 6.12  \\
    \bottomrule
  \end{tabularx}
  \caption{$L_2$-Errors of density $\rho$ for the manufactured solution at $t=1.0$ and the resultant orders of accuracy for several polynomial degrees $N$.}
  \label{tab:Conv_Sine}
\end{table}

\subsection{Scaling Tests}
\label{subsec:scaling}
Having established that the sliding mesh method produces accurate results and the interface treatment does not introduce spurious errors, we now investigate the results of the implementation and parallelization strategy described in~\secref{sec:implementation}. Favourable scaling behaviour is essential to exploit today's massively parallel hardware resources, which in turn are necessary for LES of complex applications found in industry. The baseline open source code FLEXI has been shown to scale efficiently up to over 100,000 cores~\cite{krais2020flexi,altmann2013efficient}, which allows to investigate the impact of the implemented sliding mesh method on its scaling efficiency.
To this end, scaling tests were performed on the supercomputer \hazelhen~at the High-Performance Computing Center Stuttgart~(HLRS). The Cray~XC40-system consists of 7712~nodes, each equipped with two Intel Xeon~E5-2680~v3 and 128GB of main memory.
The computational meshes for the tests are based on a cubical Cartesian mesh as shown in \figref{fig:sine_convergence} with $\ppvector{x}\in\left[0,1\right]^3$ and sliding mesh interfaces parallel to the $x_1 x_3$-plane. For the refined meshes, the amount of elements of the baseline mesh with $6\times6\times6$~elements is successively doubled in $x_1$-,$x_2$- and $x_3$-direction respectively, up to the finest mesh with $96\times48\times48$~elements.\\
The simulation is initialized with a constant freestream and the mesh velocity is set to $\ppvector{v}^g = \left[0,1,0\right]^T$. Every simulation is run exactly $100$ time steps using a five-stage Runge-Kutta method, while only the computational time excluding I/O and initialization is considered for the performance analysis. The communication partners at the sliding mesh interface change at most three times during the computation.
Starting from 1 computing node (24 cores) the amount of nodes for each mesh is then doubled until the maximum of 512~nodes (12.288~cores) is reached. However, only cases with at least 1944 DOF per core are considered, corresponding to 9 elements per core for the polynomial degree~$N=5$, which is used for the entire scaling tests. Each run is repeated five times to account for potential statistical influences like the overall network load caused by other jobs on the system.
As a consistent performance measure the performance index PID is used, which is defined as
\begin{equation}
	\mathrm{PID} = \frac{\mathrm{wall}\mhyphen\mathrm{clock}\mhyphen\mathrm{time} \cdot \mathrm{\#cores}}{\mathrm{\#DOF} \cdot \mathrm{\#time\;steps} \cdot \mathrm{\#RK}\mhyphen\mathrm{stages}} \;\;,
\end{equation}
and indicates the averaged computation walltime per spatial degree of freedom on each core for the computation of one Runge-Kutta stage.
\begin{figure}
  \includegraphics[width=\textwidth]{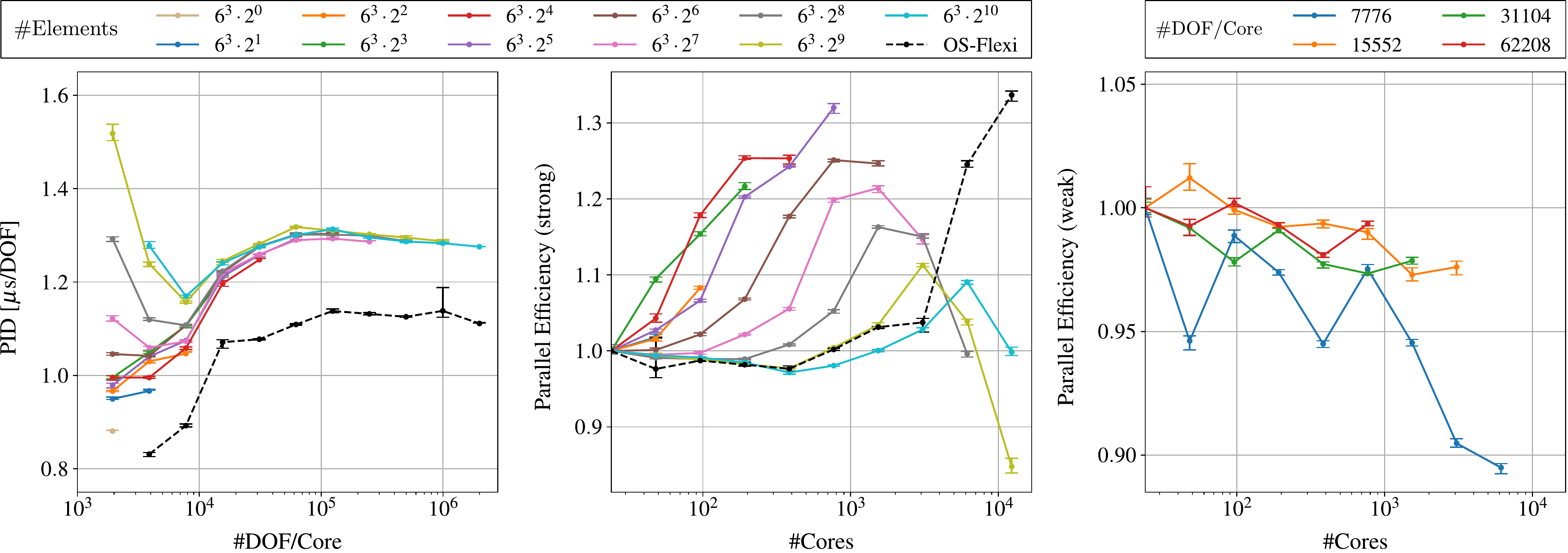}
  \caption{Strong and weak scaling of the sliding mesh implementation for different quantites of elements and computing cores on the HLRS supercomputer \hazelhen. In all plots, the mean over all five runs is given with the minimum and maximum as error bars. \textit{Left:} the performance index (PID) over the number of degrees of freedom (DOF) per core. \textit{Middle:} strong scaling as parallel efficiency over the total number of cores for the different meshes. The results for the baseline open source code (OS-Flexi) without sliding mesh for the finest grid with $6^3\cdot 2^{10}=221,184$ elements is shown dashed for comparison. \textit{Right:} weak scaling as parallel efficiency for loads ranging between 36 and 288 elements per core. }
  \label{fig:scaling}
\end{figure}
The results of the scaling tests are shown in three different plots in \figref{fig:scaling}. While the left plot shows the PID over the computational load per core, the plot in the center shows the parallel efficiency over the total amount of cores as strong scaling. The parallel efficiency is defined as ratio between the PID of the respective number of cores and the PID of the baseline simulation using the minimum of 24 cores. The results for the baseline code without sliding mesh for the finest mesh with $96\times48\times48$~elements are also plotted dashed for comparison. The right plot shows the weak scaling for four distinct loads per core.\\
For more than $10^4$ DOF per core the sliding mesh implementation shows a decrease in performance by about~$15\%$ in comparison to the baseline code. This is mainly due to the increased workload of the implemented ALE formulation and the additional overhead introduced by the sliding mesh method.
For a decreasing amount of load per core, the memory consumption per core decreases as well, which allows a greater share of data being placed in the CPU cache. The baseline code can exploit these caching effects and shows significant performance increases for small loads per core. In contrast, the overall performance of the sliding mesh code decreases significantly for low loads and large amounts of cores.
The first reason for this is the load imbalance introduced by the sliding mesh interface.
With an increasing amount of cores, the additional work at the interface is distributed to a smaller share of the cores, which leads to higher work imbalances and to a decrease of parallel efficiency.
The other reason is the increased communication effort at the sliding mesh interfaces, since the number of communication partners at the interface potentially doubles in comparison to a conforming mesh. As described in \secref{sec:implementation}, FLEXI relies on non-blocking communication which is hidden by local work. For very small loads however, the processors at the interface do not have enough local work to hide the additional communication effectively. This leads to poor scaling performance for very small loads per core.
For practical applications however, the authors have usually encountered loads of more then $10,000$ DOF per core, for which the proposed method shows excellent weak and strong scaling. The performance loss of about $15\%$ compared to the baseline code is well within acceptable limits, and could likely be further reduced by an a priori static load balancing.

\subsection{LES of the Aachen Turbine}
\label{subsec:turbine}
To demonstrate the suitability of the proposed sliding mesh implementation for large-scale simulations with the presented high-order DG scheme, it is applied to a wall-resolved implicit LES of the 1-1/2 stage \textit{Aachen turbine} test case~\cite{gallus1995,walraevens1997,volmar1998}, which is investigated extensively in literature, e.g. \cite{walraevens1998,volmar2000,yao2002,custer2012}.

\subsection*{Test Case Definition}
\label{subsec:turbine_definition}
The 1-1/2 stage Aachen turbine consists of stator vanes with modified VKI design and rotor blades with a Traupel profile \cite{utz1972} in a stator-rotor-stator setup. All blades are untwisted and the inner and outer diameter of the turbine are constant. The leading edges of the two stators are not in line, but rotated circumferentially by $3^\circ$.
Following \cite{yao2002}, the original blade count of 36-41-36 is modified to a uniform blade count, which allows to reduce the computational domain to a periodic sector with one single blade pitch per cascade. However, in contrast to \cite{yao2002}, the blade count is modified to 38-38-38 while retaining the blades' original profile geometry.
To further reduce the computational cost, the turbine is approximated by a planar cascade, which neglects any influence of the casing and the effects of the rotor's tip clearance.
The modified geometry is obtained by extruding the two-dimensional profile geometries in $x_3$-direction to a length of $6\unit{mm}$, which corresponds to $10\%$ of the rotor's chord length and approximately $10.9\%$ of the rotor's original blade span.
The remaining geometric quantities are obtained with respect to the turbine's mean radius of $r=272.5\unit{mm}$ and considering the modified blade count. For the investigated operation point with a rotational speed of $3500 \unit{rpm}$ this leads to a planar velocity of $99.88 \unit{m/s}$. The inflow Mach number is $\mathrm{Ma}\approx 0.1$ and the Reynolds number with respect to the outflow velocity and the chord length of the stator vane is $\mathrm{Re}\approx 800,000$. More details on the test case and the turbine geometry can be found in e.g. \cite{walraevens1998}.
\subsection*{Mesh Generation}
\label{subsec:turbine_meshing}
\begin{figure}
  \centering
  \includegraphics[width=\textwidth]{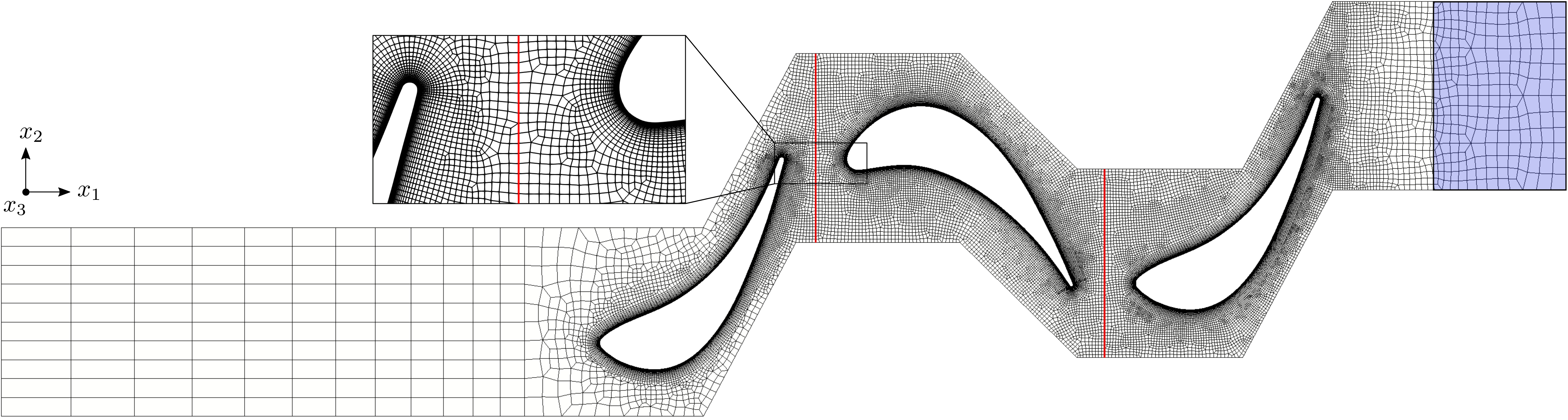}
  \caption{The cross section of the computational mesh. The sliding mesh interfaces are highlighted in red and the sponge zone at the outflow is shaded blue. The magnified section exhibits details of the transition between the structured O-grids around the blades, the equidistant mesh at the sliding mesh interface and the unstructured mesh in the remaining domain.}
  \label{fig:turbine_mesh}
\end{figure}

The mesh is based on a two-dimensional unstructured mesh with structured O-type meshes around the blades, which is extruded in $x_3$-direction with $24$ equi-spaced elements, resulting in $966,288$ hexahedral elements for the entire mesh. The sliding mesh interfaces are centered between the trailing and leading edge of consecutive blades.
The resolution at the walls is comparable to wall-resolved LES in literature \cite{gourdain2015,rodi2006} with the dimensionless wall distances for the present simulation given in \tabref{tab:wall_distance} in viscous wall units.
Our in-house high-order pre-processor HOPR~\cite{hindenlang2015} code is then used to generate a fifth-order geometry approximation of the curved blade geometry and an RBF method is used to expand the curving into the surrounding volume, as described more detailed in \cite{krais2020flexi}.
\begin{table}
  \centering
  \setlength{\tabcolsep}{8pt}
  \begin{tabularx}{\textwidth}{llXcccXcccXccc}
    \toprule
             &                    &&        & $x^+$  &       &&       & $y^+$ &       &&        & $z^+$  &       \\
             &                    && max    & mean   & std   && max   & mean  & std   && max    & mean   & std   \\
    \midrule
    Rotor    & $\mathcal{T}=0.0$  && $36.6$ & $23.8$ & $8.1$ && $2.2$ & $1.4$ & $0.4$ && $26.6$ & $16.3$ & $4.4$ \\
             & $\mathcal{T}=0.5$  && $38.2$ & $24.2$ & $8.1$ && $2.5$ & $1.4$ & $0.4$ && $29.4$ & $16.7$ & $4.5$ \\
    \norule
    Stator 2 & $\mathcal{T}=0.0$  && $42.8$ & $24.1$ & $9.5$ && $2.2$ & $1.3$ & $0.4$ && $29.1$ & $17.6$ & $5.1$ \\
             & $\mathcal{T}=0.5$  && $44.5$ & $25.4$ & $9.3$ && $2.2$ & $1.4$ & $0.4$ && $29.2$ & $18.7$ & $5.3$ \\
    \bottomrule
  \end{tabularx}
  \caption{The dimensionless wall distances for the wall-nearest grid cells of the rotor blade and the second stator vane in viscous wall units. The wall distances are normalized with the factor $(N+1)$ to account for the multiple DOF in each direction inside a DG element. Given are the respective maximum, mean and the standard deviation of the phase-averaged wall distances at two distinct phase angles $\mathcal{T}=0.0$ and $\mathcal{T}=0.5$.}
  \label{tab:wall_distance}
\end{table}\\
The boundary conditions are set periodic in  $x_2$- and $x_3$-direction and the blade walls are modelled as adiabatic no-slip walls. The measured inflow state at the mean diameter of the turbine from the test case's experimental data is imposed as Dirichlet boundary condition at the inflow position. Similarly, the measured pressure behind the second stator is used to impose a pressure outflow condition~\cite{carlson2011}. Both states are given in \tabref{tab:inflow_outflow_turbine}.
A sponge zone with ramping function is employed before the outflow to avoid spurious reflections at the outflow, as depicted in \figref{fig:turbine_mesh}. More details on the used sponge zone and ramping function can be found in \cite{flad2014}.

\begin{table}
  \centering
  \setlength{\tabcolsep}{14pt}
  \begin{tabularx}{\textwidth}{Xccccc}
    \toprule
            & $\rho$          & $p$            & $v_1$        & $v_2$         & $v_3    $ \\
    \midrule
    Inflow  & $1.7765 \unit{kg/m^3}$ & $157305.88 \unit{Pa}$ & $37.200 \unit{m/s}$ & $-2.010  \unit{m/s}$ & $0.0 \unit{m/s}$  \\
    Outflow & $1.3651 \unit{kg/m^3}$ & $110357.08 \unit{Pa}$ & $55.992 \unit{m/s}$ & $160.019 \unit{m/s}$ & $0.0 \unit{m/s}$  \\
    \bottomrule
  \end{tabularx}
  \caption{The measured inflow and outflow conditions at the mean diameter of the turbine from the test case's experimental data. The inflow was measured $143 \unit{mm}$ in front of the leading edge of the first stator and the outflow state is given $8.8 \unit{mm}$ behind the trailing edge of the second stator. The velocity $v_3$ is set to zero at the inflow and outflow on account of the planar cascade assumption.}
  \label{tab:inflow_outflow_turbine}
\end{table}
\subsection*{Computational Setup}
\label{subsec:turbine_computational_requirements}
For the simulation, a sixth-order scheme is employed, which results in approximately $208$ million spatial degrees of freedom. A fourth-order Runge-Kutta method by Niegemann et al. \cite{niegemann2012} is used, since its optimized stability region allows for larger time steps. The specific gas constant is set to $R=287.058\unit{\frac{J}{kg\cdot K}}$, the viscosity to $\mu=1.8\cdot 10^{-5} \unit{\frac{kg}{m\cdot s}}$, the ratio of specific heats to $\gamma=1.4$ and the Prandtl number is chosen as $Pr=0.72$. The LES is filtered implicitly by the discretization with the discretization error also acting as an implicit subgrid scale model, following e.g.~\cite{beck2014}.\\
The computation was carried out on \hazelhen\ at the HLRS with a varying amount of up to $4800$ cores, resulting in a minimum of $41,300$ DOF per core. The PID was approx. $1.3\unit{\mu s}$ per DOF for all considered cases and shows good agreement with the results of the scaling test. After attaining a quasi-periodic solution, $10$ periods of the turbine flow were computed with a computational cost of about $27,000$ CPU-hours per period.

\subsection*{Results}
\label{subsec:turbine_results}
\begin{figure}[htbp!]
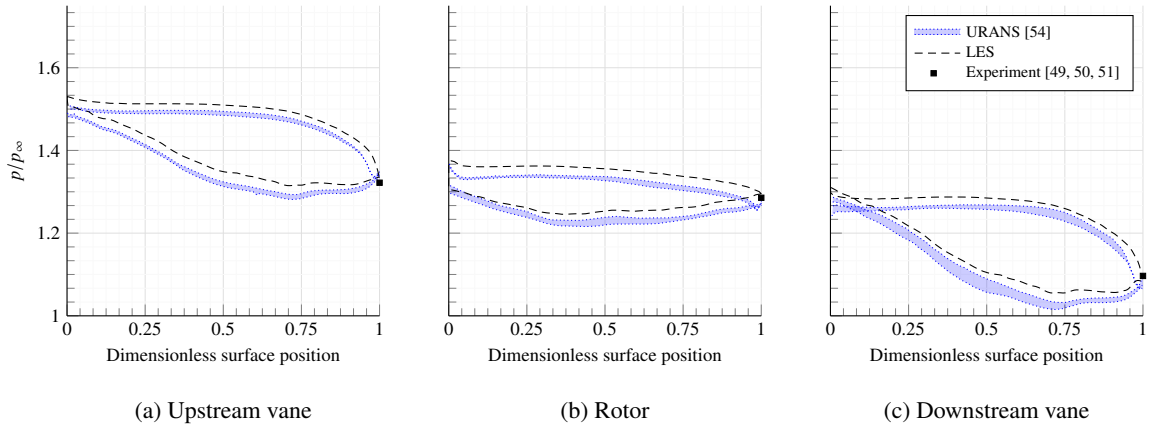
%
	\centering
  \begin{subfigure}[t]{.25\linewidth}%
    \includegraphics[width=\linewidth]{tikz/fig_timeavg_v1_cp.tikz}%
    \caption{Upstream vane}
    \label{fig:timeavg_pressure_v1}
  \end{subfigure}
  \hspace{.05\linewidth}%
  \begin{subfigure}[t]{.25\linewidth}%
    \includegraphics[width=\linewidth]{tikz/fig_timeavg_r1_cp.tikz}%
    \caption{Rotor}
    \label{fig:timeavg_pressure_r1}
  \end{subfigure}
  \hspace{.05\linewidth}%
  \begin{subfigure}[t]{.25\linewidth}%
    \includegraphics[width=\linewidth]{tikz/fig_timeavg_v2_cp.tikz}%
    \caption{Downstream vane}
    \label{fig:timeavg:pressure:v2}
  \end{subfigure}
  \caption{Comparison of static pressure. The results of the LES (black dashed) are compared with the unsteady pressure envelopes by Yao et al.~\cite{yao2002} (blue) and the pressure at the trailing edges from experimental data \cite{gallus1995,walraevens1997,volmar1998} (black squares).}
	\label{fig:timeavg_pressure}
\end{figure}

\begin{figure}[htb]
  \newcommand\plotwidth{0.49\textwidth}
  \newcommand\plotspace{0.009\textwidth}

  \begin{subfigure}{\plotwidth}
    \caption*{$\mathcal{T}=0$ : Configuration with \textbf{low} rotor lift}
    \raggedright
    \includegraphics[width=\textwidth]{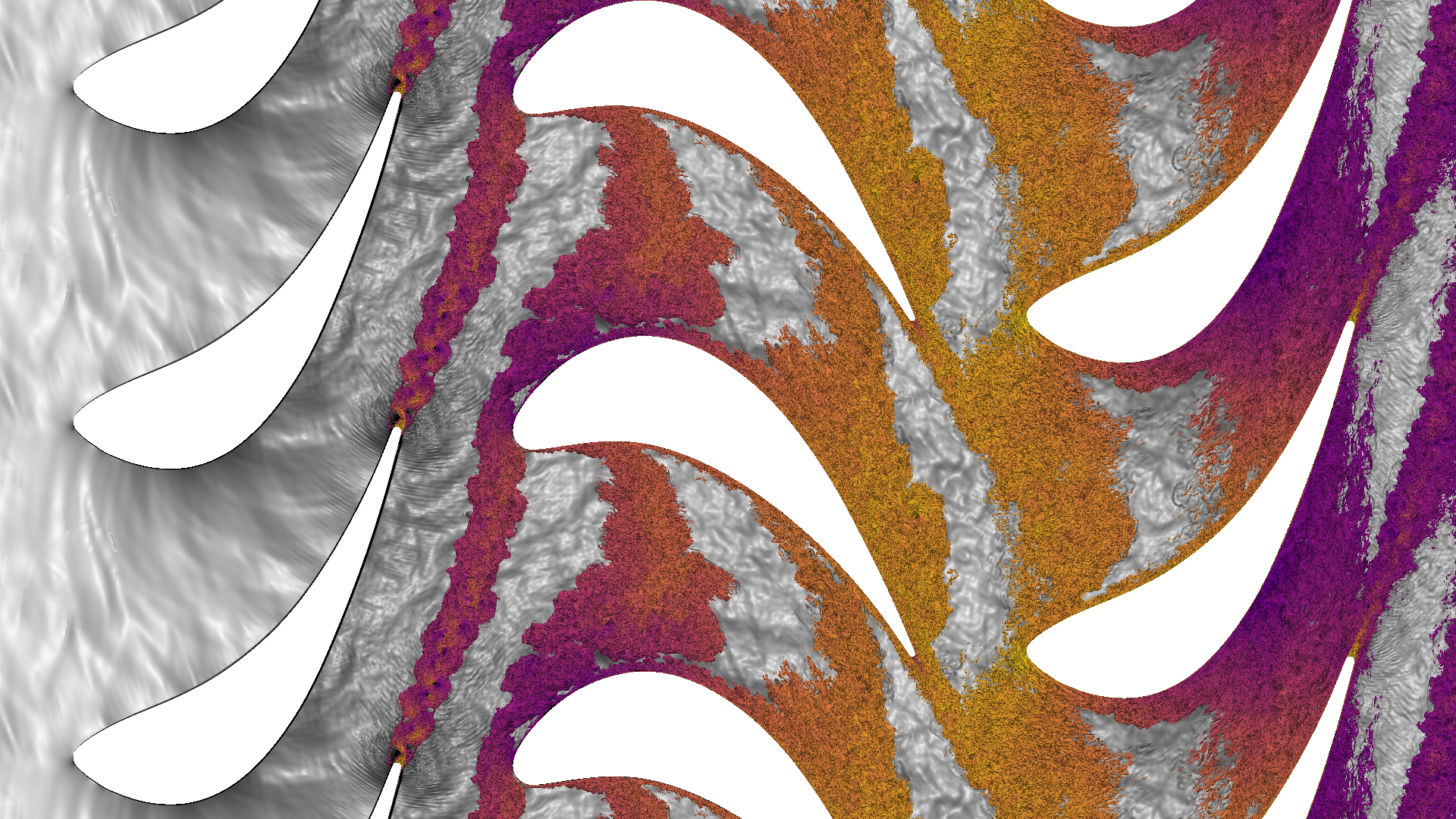}
  \end{subfigure}
  \hspace{\plotspace}
  \begin{subfigure}{\plotwidth}
    \caption*{$\mathcal{T}=\frac{1}{2}$ : Configuration with \textbf{high} rotor lift}
    \raggedleft
    \includegraphics[width=\textwidth]{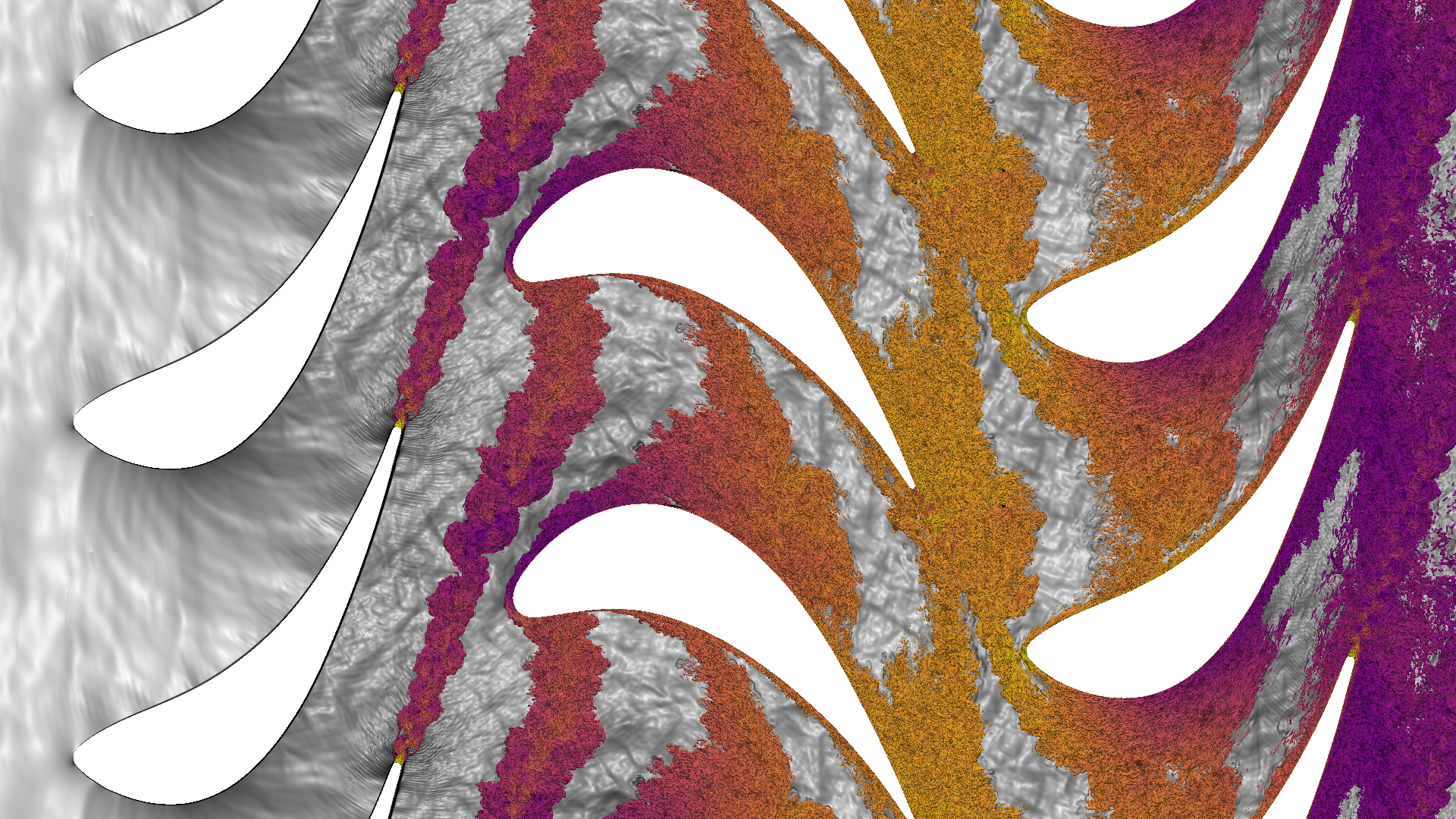}
  \end{subfigure}\\
  \par\vspace*{\plotspace}
  \begin{subfigure}{\plotwidth}
    \raggedright
    \includegraphics[width=\textwidth]{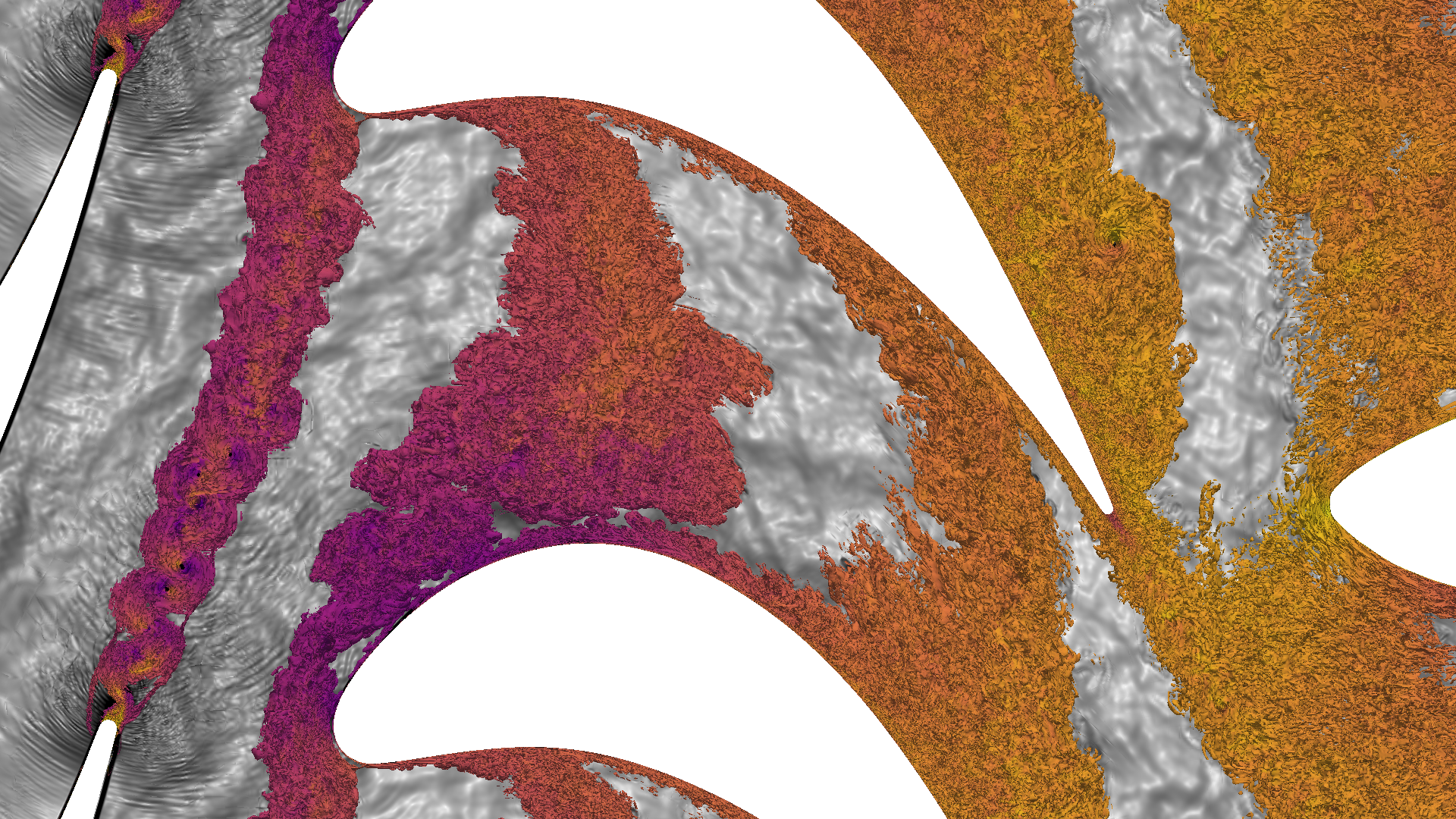}
  \end{subfigure}
  \hspace{\plotspace}
  \begin{subfigure}{\plotwidth}
    \raggedleft
    \includegraphics[width=\textwidth]{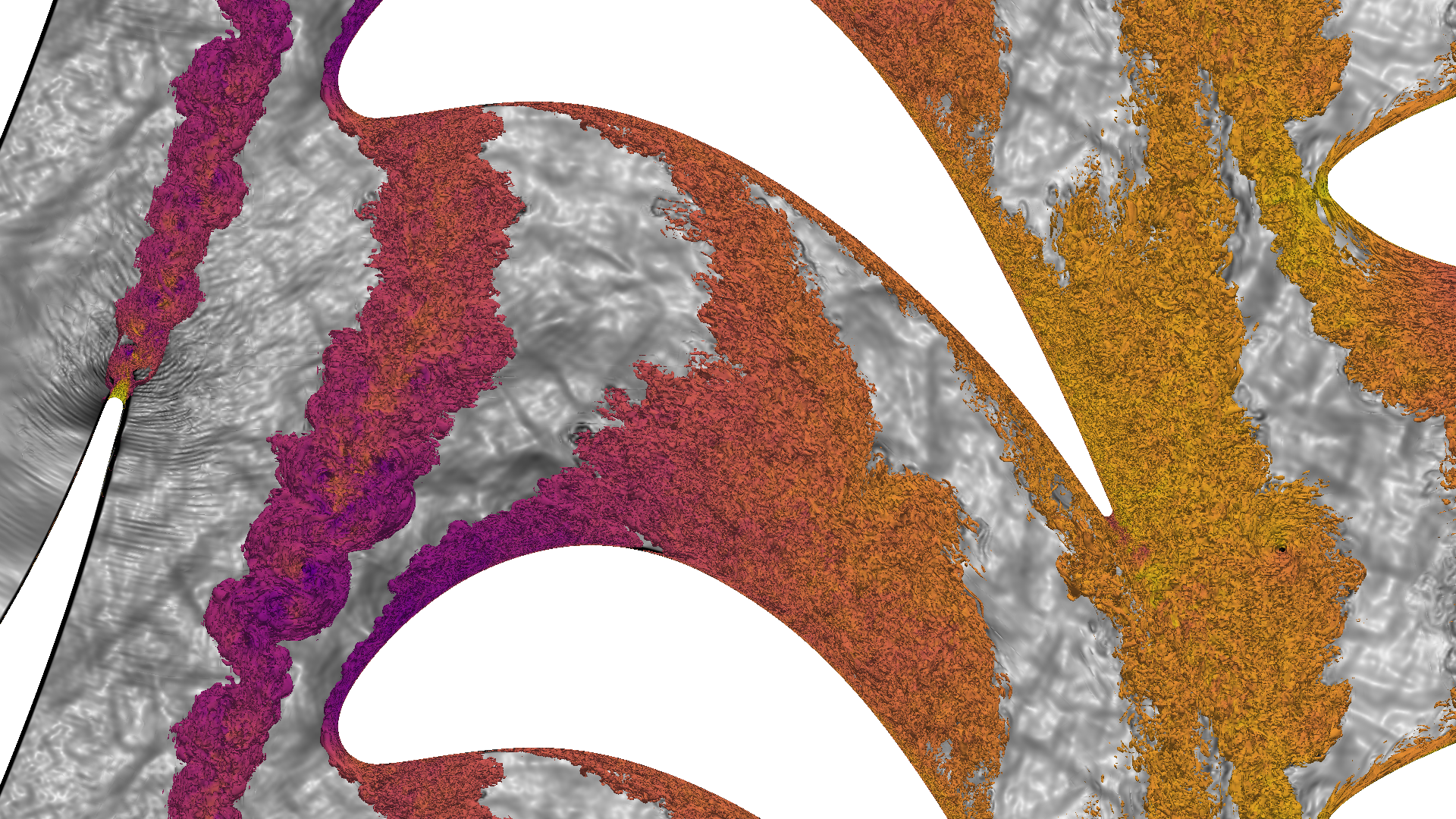}
  \end{subfigure}\\
  \par\vspace*{\plotspace}
  \begin{subfigure}{\textwidth}
    \centering
    \includegraphics[width=0.4\textwidth]{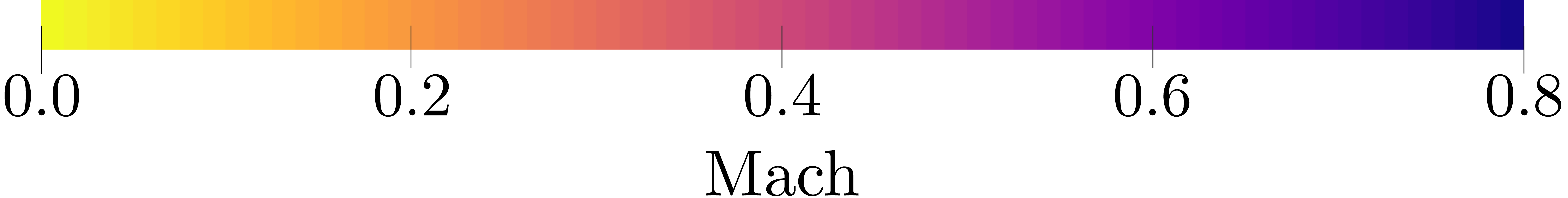}
  \end{subfigure}
  \caption{Instantaneous flow field visualized with iso-surfaces of the $\lambda_2$-criterion \cite{jeong1995} colored by the Mach number in front of pseudo-schlieren computed at $x_3=0\unit{mm}$ with an offset of half a period between the left and right figures. The lower figures show a close-up of the passage between two rotor blades at the respective time instants.}
  \label{fig:turbine_schlieren}

  \let\plotwidth\undefined
  \let\plotspace\undefined
\end{figure}

\begin{figure}[htbp!]
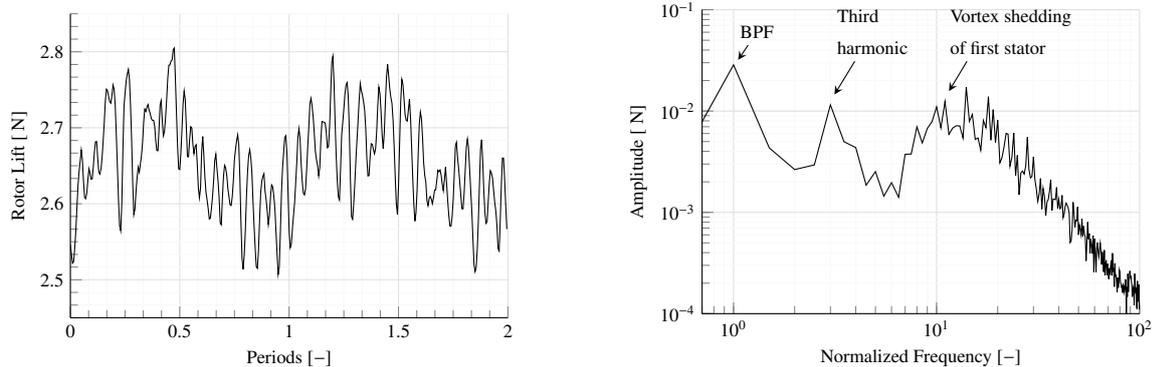
%
	\centering
  \begin{subfigure}[t]{.35\linewidth}%
    \includegraphics[width=\linewidth]{tikz/fig_lift_time_domain.tikz}%
  \end{subfigure}
  \hspace{.15\linewidth}%
  \begin{subfigure}[t]{.35\linewidth}%
    \includegraphics[width=\linewidth]{tikz/fig_lift_freq_domain.tikz}%
  \end{subfigure}
  \caption{The lift force acting on the rotor blade. The results are averaged by dividing the obtained lift force into intervals of two rotor periods. These intervals are then averaged to obtain the temporal evolution of the lift force on the left. Shown on the right is the lift force in the frequency domain as average of the discrete Fourier transforms of the individual time intervals. Highlighted are the blade passing frequency (BPF), its third harmonic and the frequency of the first stator vane's vortex shedding. The frequency is normalized with the BPF.}
  \label{fig:turbine_body_forces}
\end{figure}

The time-averaged surface pressure of the vanes and the rotor blade are given in \figref{fig:timeavg_pressure} together with the results of Yao et al.~\cite{yao2002} for comparison, who conducted an URANS simulation of a three-dimensional blade passage based on a 36-36-36 blade configuration. The results show good qualitative agreement even though the LES predicts an overall higher pressure level than the URANS simulation. Furthermore, the LES shows excellent agreement with the available experimental data at the blades' trailing edges.
\\
The instantaneous flow fields for two distinct phase angles $\mathcal{T}=0$ and $\mathcal{T}=\frac{1}{2}$ are given in \figref{fig:turbine_schlieren}, with $\mathcal{T}$ as the relative rotor position and $\mathcal{T}=1$ corresponding to one completed period of the rotor.
The flow around the first stator vane remains laminar with transition just before the trailing edge on the suction side, due to the strong favourable pressure gradient. The induced vortex shedding causes aeroacoustic noise, which is also convected upstream, as exhibited by the numerical pseudo-schlieren.
The vane's wake impinges on the rotor and wraps around the rotor's leading edge. The wake's axis is then rotated counter-clockwise before it is passively advected by the freestream, as detailed in \cite{wu2001}. The wake acts as perturbation in the rotor passage, creating unsteady pressure distributions on the rotor surface, e.g. \cite{coull2011}.
To quantify these effects and their impact on the rotor, the averaged lift force acting on the rotor blade as well as its spectrum is given in~\figref{fig:turbine_body_forces}. The data is obtained by evaluating the lift force and its Fourier transform on intervals with 2 periods each and averaging the obtained results.
The minimum lift force is reached at around $\mathcal{T}\approx 0.8$ when the wake contacts the suction side of the rotor. In contrast, the rotor lift force increases as the point of impact shifts towards the rotor's pressure side, reaching the maximum lift force at $\mathcal{T}\approx 0.3$.
Interestingly, only the third harmonic of the blade passing frequency (BPF) is distinguishable, while the second and fourth harmonic show no considerable contribution to the lift force.
The increasing amplitudes for frequencies at around 10 BPFs are caused by the vortex shedding of the first stator vane, as their frequencies conincide.
\\
As this initial analysis of the resulting flow field reveals, a number of non-linear interactions are triggered by the complex stator/rotor interactions, which warrant further detailed analysis. Of particular interest will be the boundary layer state and its interaction with the wakes, as well as the resulting load transients. As the focus of this paper is on the methodology and the performance of the described sliding mesh implementation, a more throughout analysis of the flow physics will be subject of future research. However, this test case here already highlights the potential of the presented high-order sliding mesh method for scale-resolving simulations in turbomachinery and related applications.

\section{Conclusion}
\label{sec:conclusion}
In this work,  we have proposed an efficient implementation and parallelization strategy  of a mortar-based sliding mesh method for high-order discontinuous Galerkin methods. The method retains the high-order accuracy of the DG method as well as the excellent strong and weak scaling properties of the baseline code, and is thus well-prepared to tackle problems at an industrial scale. \\
The challenge in designing a parallelization strategy lies in the dynamic communication structure. At the heart of our proposed approach lies the avoidance of additional global communication as well as the passing of metadata. Instead, only essential solution data is communicated, while process-local mapping arrays handle the identification of communication partners at each timestep. The presented, globally unique sorting strategy keeps the message data contiguous in memory and avoids local rearranging of the data on arrival. After validating the accuracy and convergence property of our approach, we demonstrate that due to the careful design of the parallelization scheme, the sliding mesh implementation achieves excellent strong and weak scaling. We note that the optimum load per core is shifted slightly towards higher loads, and that the performance deteriorates for very low loads per core. This is to be expected and will be improved in the future with additional load balancing. Compared to the baseline scheme, a performance loss of only about $15\%$ is incurred by the novel method, which is acceptable and allows us to conduct large scale simulations with sliding mesh interfaces on the available supercomputers. We present an example of such an application for the case of a turbine flow with stator-rotor-stator interaction. To the authors' best knowledge, this is the first time a high-order sliding mesh method for DG has been applied to large scale problems of industrial relevance.\\
In the future, we plan to apply this framework to a range of interesting cases, where high unsteadiness and non-linear interactions pose challenging problems for traditional models like the RANS equations. A typical example of such applications can be found in turbomachinery components, where our simulation framework can contribute to the understanding of complex 3D flows. Beyond the pure fluid phase however, coupling the sliding mesh approach with a particle tracking method as developed in~\cite{beck2019towards} will establish simulation capabilities that can investigate particle-laden flows in rotating geometry, which are rarely tractable with an experimental approach.

\section*{Acknowledgment} %
The research presented in this paper was funded by Deutsche Forschungsgemeinschaft (DFG, German Research Foundation) under Germany's Excellence Strategy - EXC 2075 - 390740016 and by Friedrich und Elisabeth Boysen-Stiftung as part of the project BOY-143. 

The authors gratefully acknowledge the support and the computing time on "Hazel Hen" provided by the HLRS through the project "hpcdg".

The measurements on the test case "Aachen Turbine" were carried out at the Institute of Jet Propulsion and Turbomachinery at RWTH Aachen
University, Germany.

\appendix

\section{Illustrative Mortar Sorting Example}
\label{sec:example}

Here, the construction of the mapping arrays $\mrank$, $\sortedArray$ and $\mapping$ laid out in \secref{sec:implementation_sorting} is illustrated with an example. 

\begin{figure}[h]
  \centering
  \begin{subfigure}[c]{0.3\textwidth}
    \centering
    \includegraphics[width=0.8\textwidth,trim=0mm 0mm 20mm 0mm,clip=true]{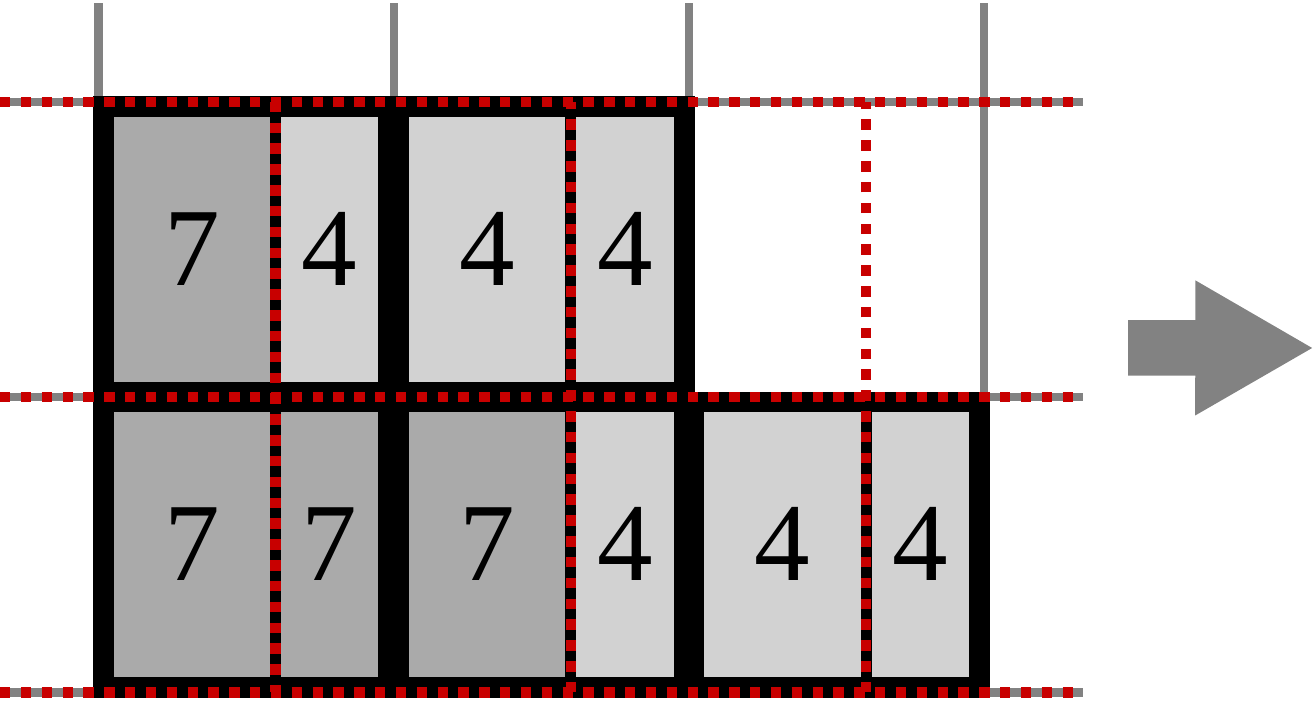}
    \subcaption{$\mov{\rank}$}
  \end{subfigure}
  \begin{subfigure}[c]{0.3\textwidth}
    \centering
    \includegraphics[width=0.8\textwidth,trim=0mm 0mm 20mm 0mm,clip=true]{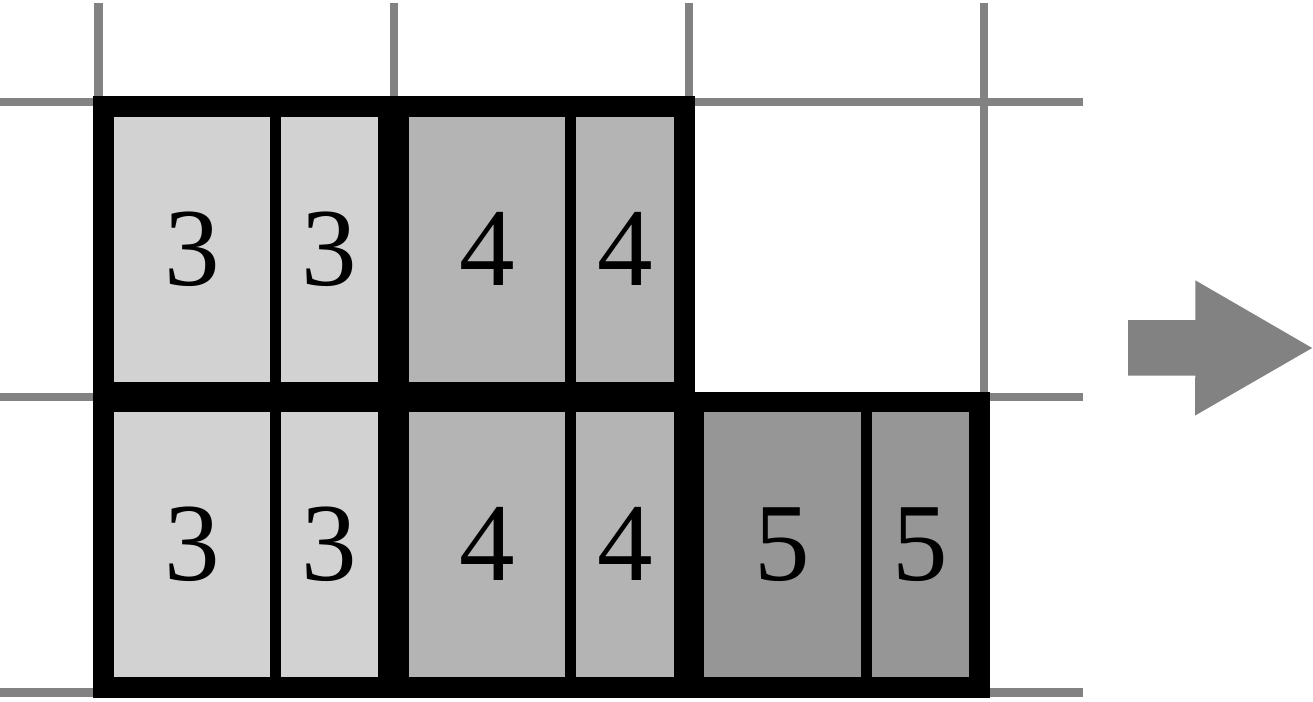}
    \subcaption{$\sipa$}
  \end{subfigure}
  \begin{subfigure}[c]{0.3\textwidth}
    \centering
    \includegraphics[width=0.8\textwidth,trim=0mm 0mm 20mm 0mm,clip=true]{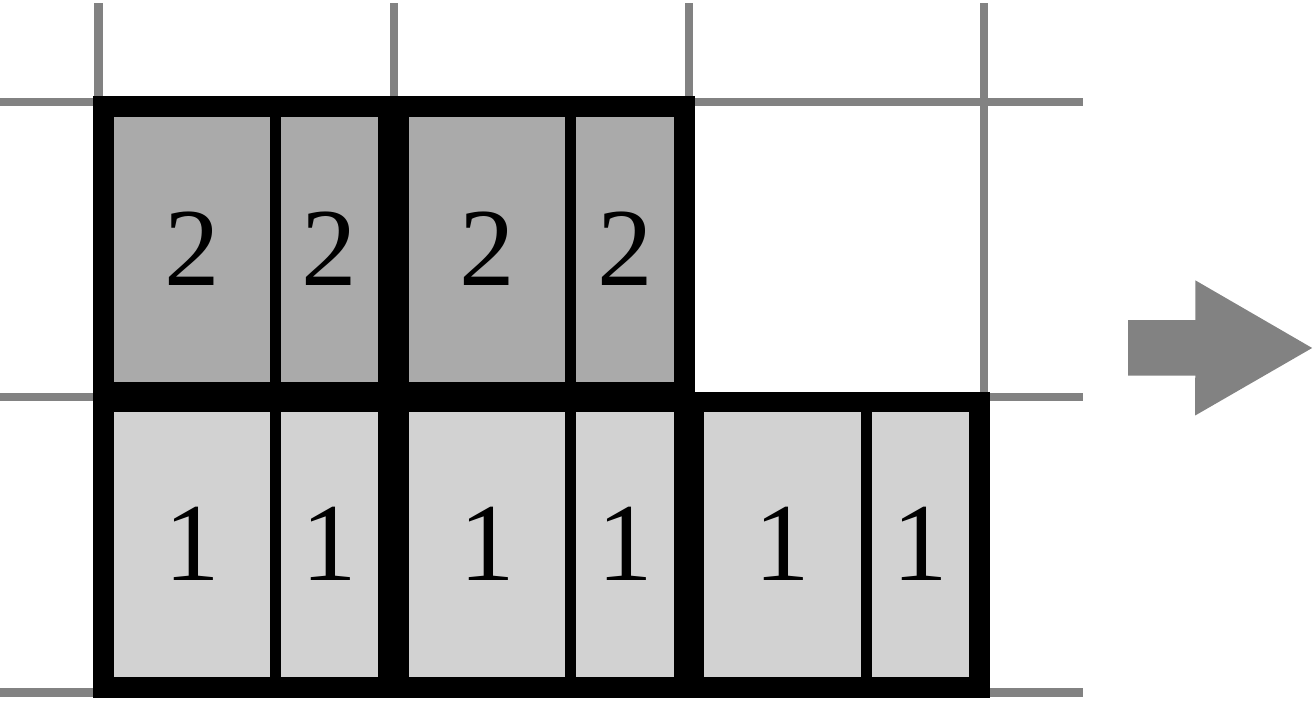}
    \subcaption{$\sipe$}
  \end{subfigure}\\
  \vspace*{0.3cm}
  \begin{subfigure}[c]{0.36\textwidth}
    \centering
    \includegraphics[width=0.64\textwidth,trim=0mm 0mm 20mm 0mm,clip=true]{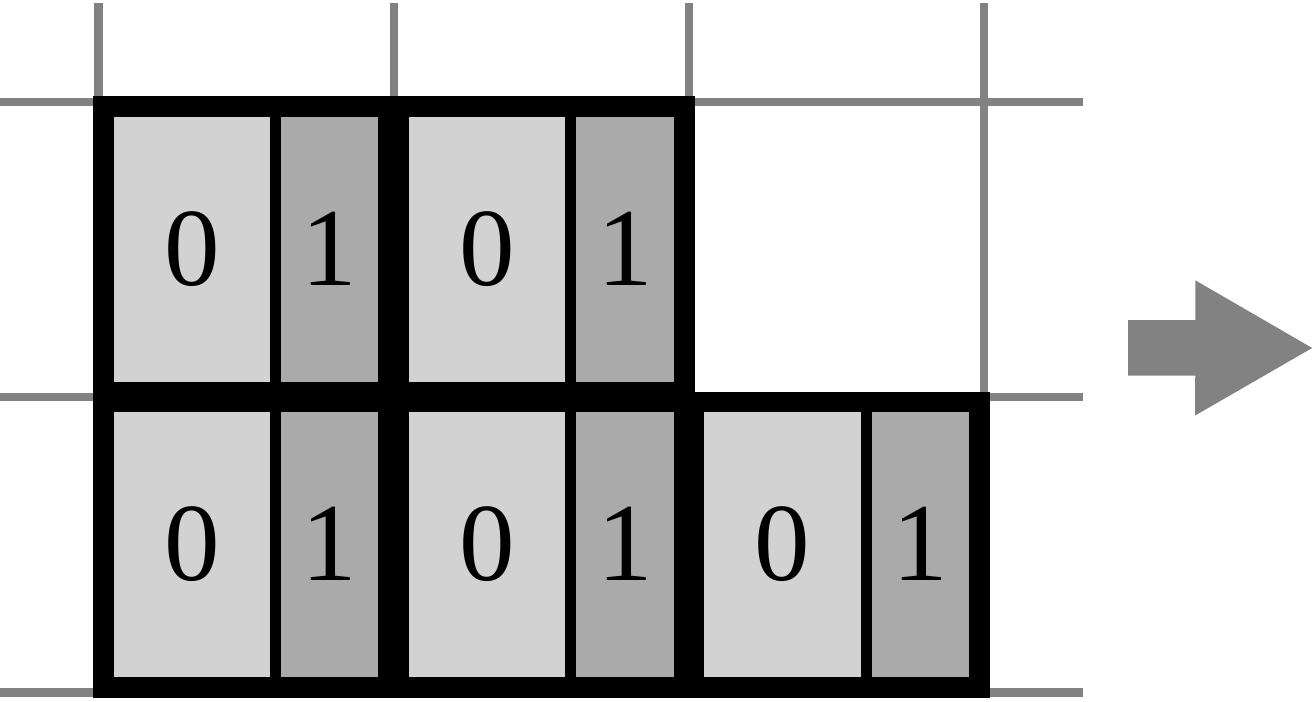}
    \subcaption{$\im$}
  \end{subfigure}\begin{subfigure}[c]{0.36\textwidth}
    \centering
    \includegraphics[width=0.64\textwidth,trim=0mm 0mm 20mm 0mm,clip=true]{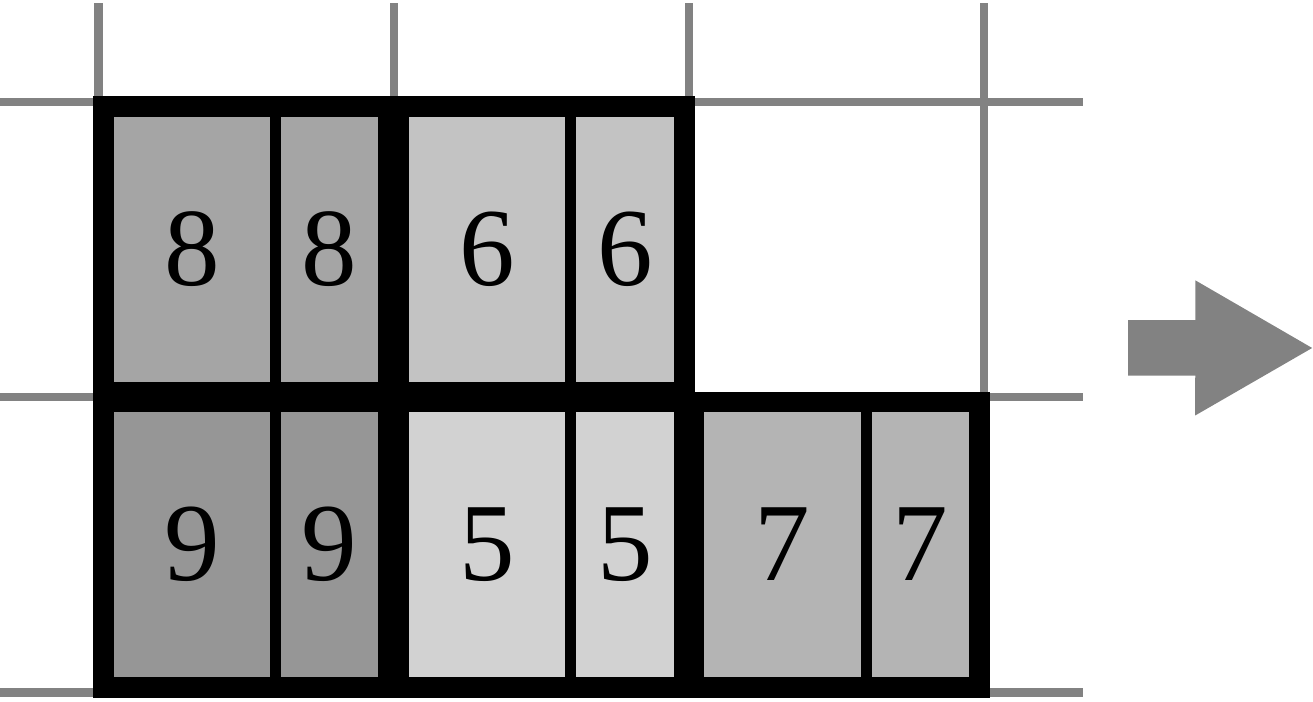}
    \subcaption{$\flexiSideID$}
  \end{subfigure}
  \caption{Example indices of sliding mesh sides of a process in the static subdomain. The movement of the neighbouring moving domain is from left to right.\label{fig:example_setup}} 
\end{figure}

The example setup is shown in \figref{fig:example_setup} and described in the following: The considered process on the static subdomain handles five sliding mesh sides (thick black lines). Each of them contains two sliding mesh mortars (fine black lines). The moving subdomain moves from left to right. At the considered time instant, the adjacent sides in the moving subdomain are handled by the processes with ranks 4 and 7, their indices are given by $\mrank$ (Sub-Figure a). The moving sides are outlined with red dotted lines. They are not aligned with the static sides. Furthermore, in this example, the five sliding mesh sides handled by the considered static process have global static indices $\sipa$ ranging from 3 to 5 (Sub-Figure b) and $\sipe$ ranging from 1 to 2 (Sub-Figure c). Within each side, the mortar sub-index is ascending along the direction of movement as stated in \eqref{eq:subim} (Sub-Figure d). The side indices assigned to each side by FLEXI are process-local and arbitrary, but continuous (Sub-Figure e).

The ranks of{\it~all} processes on the moving domain adjacent to the interface interface are stored in the array $\mrank$, which is sent to every static process adjacent to the interface. The column index of $\mrank$ is $\mipa$ and the row index is $\sipe$. For each static side, the static process knows the static global indices $\sipa$ and $\sipe$. It calculates the column indices $\mipa$ from $\sipa$ using \eqref{eq:indices} and reads the ranks shown in Sub-Figure a from the according entries of $\mrank$, which are
\begin{equation}
  \mrank = 
      \begin{pmatrix}
        \cdots & 7 & 7 & 4 & 4 & \cdots \\
        \cdots & 7 & 4 & 4 & \ddots & \cdots \\
        \iddots  & \vdots  & \vdots & \vdots & \vdots & \ddots 
      \end{pmatrix}.
\end{equation}
Note that compared to the figure, the orientation of $\sipe$ is flipped. 

The indices of all mortars handled by one process are stored in $\sortedArray$. The columns of $\sortedArray$ contain the different mortars, the rows contain the different types of indices. For the considered example, the array $\sortedArray$ after sorting looks as follows: 
\begin{equation}
   \sortedArray = 
      \begin{pmatrix}
        4 & 4 & 4 & 4 & 4 & 4 & 7 & 7 & 7 & 7 \\
        3 & 4 & 4 & 4 & 5 & 5 & 3 & 3 & 3 & 4 \\
        2 & 1 & 2 & 2 & 1 & 1 & 1 & 1 & 2 & 1 \\
        1 & 1 & 0 & 1 & 0 & 1 & 0 & 1 & 0 & 0 \\
        8 & 5 & 6 & 6 & 7 & 7 & 9 & 9 & 8 & 5
      \end{pmatrix}\,\,\begin{matrix*}[l]
        \mov{\rank} \\
        \sipa \\
        \sipe \\
        \im \\
        \flexiSideID
      \end{matrix*}
\end{equation}
This is the result of sorting the mortars (that is the columns) hierarchically by the entries of the upper four rows: The upper row  $\mrank$ determines the highest (outermost) sorting criterion and the fourth row $\im$ the lowest. The entries of the last row $\flexiSideID$ do not contain a sorting criterion and are transported passively. 

\begin{figure}[h]
    \centering
    \includegraphics[width=0.24\textwidth,trim=0mm 0mm 20mm 0mm,clip=true]{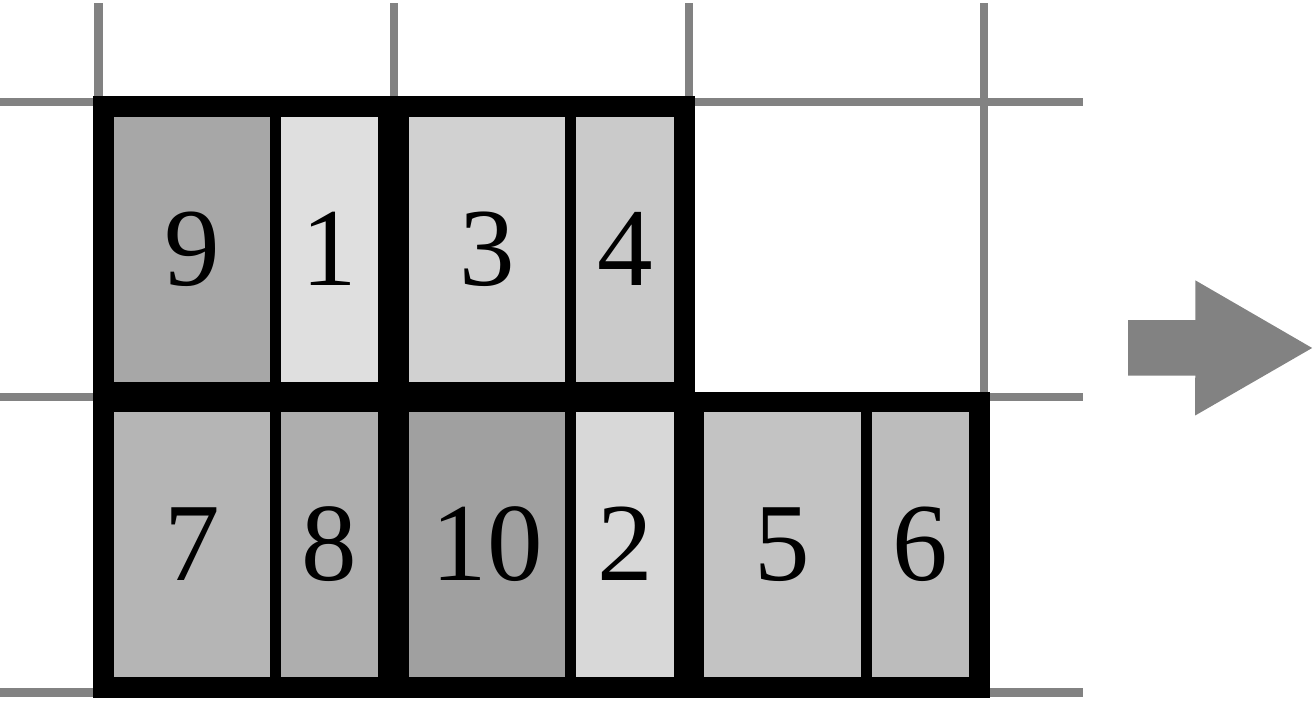}
  \caption{Mortar index $\iMortar$ as a result of sorting the columns of $\sortedArray$. \label{fig:example_imortar}}
\end{figure}

The columns of $\sortedArray$ are now indexed from left to right, that is from 1 to 10. This index is called $\iMortar$. It is depicted in \figref{fig:example_imortar}. For example, the upper left mortar belongs to the ninth column of $\sortedArray$. As can be verified in the different sub-figures of \figref{fig:example_setup}, the column entries match the indices of the upper left mortar. 

While the{\it~upper four} rows of $\sortedArray$ are needed to create a globally unique sorting, the{\it~lower two} rows are needed locally for the mapping from local sides to global sorting. To this end, the mapping array $\mapping$ is filled. It represents the inverse mapping of the last two rows of $\sortedArray$. Its rows correspond to $\im$ (ranging from 0 to 1) and its columns to $\flexiSideID$. Note that the column index does not necessarily start at 1, but in our example ranges from 5 to 9. The entries of $\mapping$ are the mortar indices $\iMortar$, such that in the considered example, 
\begin{equation}
   \mapping = 
      \begin{pmatrix}
         10 & 3 & 5 & 9 & 7 \\
         2  & 4 & 6 & 1 & 8 \\
      \end{pmatrix}.
\end{equation}
As a verifying example, the upper left entry of $\mapping$ with row index 0 and column index 5 has the value $\iMortar=10$, while in the tenth column of $\sortedArray$, $\im=0$ and $\flexiSideID=5$. 

As described in \secref{sec:implementation_sorting}, the array $\mapping$ is now used to store the solution and Flux on the mortars in their own arrays $U_\text{primary}^\text{sm}$, $U_\text{replica}^\text{sm}$, $F_\text{primary}^\text{sm}$ and $F_\text{replica}^\text{sm}$ using only one index $\iMortar$ to distinguish between the mortars. Its has the desired properties described in \secref{sec:implementation_sorting} in that contiguous chunks of data are sent and received during communication and the order of the data is globally defined and thus identical for sending and receiving process. To this end, a similar sorting procedure is carried out on the moving subdomain, too.

\bibliographystyle{elsarticle-num}
\bibliography{bibliography}

\end{document}